\documentclass[12pt]{article}

\usepackage{graphicx}
\usepackage{subfigure}
\usepackage{amsmath}
\usepackage{enumerate}
\usepackage{bm}
\usepackage{natbib}
\usepackage[hyphens]{url}
\usepackage{mathrsfs}
\usepackage{threeparttable}
\usepackage{hyperref}
\usepackage{multirow}

\newcommand{\eg}{e.g.,}
\newcommand{\etal}{et al.}
\newcommand{\etc}{etc.}
\newcommand{\ie}{i.e.,}

\begin{document} 

\setlength{\footskip}{0pt} 
\begin{center}
\vspace*{\fill}
\textbf{\large Ejecta cloud from the AIDA space project kinetic impact on the secondary of a binary asteroid: II. fates and evolutionary dependencies}\\ 
\vfill
\textbf{Yang Yu}$^\star$\\ 
\textbf{Patrick Michel}$^\dag$\\ 

\bigskip

$^\star$Beihang University\\
100191 Beijing, China\\

$^\dag$Laboratoire Lagrange, Universit\'e C\^ote d'Azur,\\
Observatoire de la C\^ote d'Azur, CNRS\\
06304 Nice, France\\

\bigskip
Printed \today\\

\bigskip
Submitted to \textit{Icarus}\\
\vfill
66 manuscript pages\\
8 figures including 5 in color (online version only)
\vfill
\end{center}

\newpage

\begin{flushleft}
Proposed running page head: Dynamics of Ejecta Cloud in a Binary Asteroid System\\
\bigskip
Please address all editorial correspondence and proofs to: Yang Yu\\
\bigskip
Yang Yu\\
No. 37 Xueyuan Rd., Haidian District\\
Beihang University, 100191 Beijing, China\\
Tel: +86 134 6676 6439\\
E-mail: \texttt{yuyang.thu@gmail.com}
\end{flushleft}

\newpage
\tableofcontents
\newpage
\listoffigures
\newpage
\listoftables

\newpage
\section*{Abstract}

This paper presents a quantitative study of the evolution of the ejecta cloud released from a hypervelocity impact on a binary asteroid. The Asteroid Impact \& Deflection Assessment (AIDA) mission project in collaboration between NASA and ESA aims to perform an asteroid deflection demonstration, using a half-ton projectile that will perform a hypervelocity impact on the surface of the secondary of the binary near-Earth asteroid ($65803$) Didymos, called hereafter Didymoon. We performed numerical simulations of the post-impact dynamics of the ejecta cloud in the framework of the current mission scenario of AIDA. Our analysis relies  on a classification of the orbits as a function of the ejecta fates, \eg\ a collision with  one of the binary components or the escape from the region of influence of the system. A grid search of launching sites of ejecta was defined over the globe of Didymoon, and considering a wide range of possible ejection speeds, we determined the dependency of ejecta fate on launching sites (projectile impact sites) and speeds. This range allows us to track all the complex cases that include different types of dynamical fates. The results reveal the detailed proportions of the ejecta that are either orbiting, escaping or re-accreting on the primary/secondary at the end of the considered timescale, as a function of the ejection speed, which allows us to explore the global characteristics of the ejecta dynamical fates. Two major mechanisms are found to be working broadly during the post-ejection evolution of the ejecta cloud: 1) ejecta on mean motion resonance orbits with Didymoon produce long-term quasi-periodic showers onto Didymoon over at least a couple of weeks after the projectile impact, 2) ejecta on non-resonant orbits produce a rapid and high re-accretion flux. This rapid and high flux occurs just once because ejecta on such orbits leave the system unless they experience a collision during their first encounter. For both mechanisms, swing-bys of Didymoon are found to occur. These swing-bys are a source of chaotic motion because the outcome of the swing-by is extremely sensitive to the ejecta initial conditions. Moreover, for all ejecta speeds, a zone free of ejecta is noticed to emerge around the mid-latitude zone of the celestial sphere about two months after the projectile impact. Also, the extent of this zone depends on the ejecta speed. For the second part of this study, we performed full-scale simulations of the ejecta cloud released from $6$ hypothetical impact sites. To define the initial conditions of the ejecta based on cratering scaling laws, we considered two kinds of material composing Didymoon's subsurface and then combined a power-law size distribution of the ejecta with an ejection speed distribution. We find that the ejecta cloud evolution can be divided in two periods. It starts with a first violent period ($<10$ hr) with fast re-accretion or ejection of the ejecta from the system. A second period is found to be more sensitive to the launching site than the first one. During this second period, ejecta will either re-accrete or being ejected from the system, depending both on their sizes and on their average survival time in close proximity of the binary components. There is thus a size-sorting effect dictated by the solar radiation pressure, which proves to be efficient to move out of the system the dust-size ejecta ($<1$ mm) for all considered launching sites and material types. On the other hand, the larger ejecta, being less or not affected by the solar radiation pressure, can survive longer in the system.

\begin{description}
\item{\textbf{Keywords}: Asteroid, dynamics; Collisional physics; Impact processes; Debris disks; Accretion.}
\end{description}

\newpage
\section{Introduction} \label{s:sec1_intro}

An ejecta cloud is one of the most expected outcome of an asteroid impact mission, such as the AIDA space project in collaboration between NASA and ESA. If funded for launch, AIDA will be the first kinetic impact deflection test. Its objective is to characterise the near-Earth binary asteroid ($65803$) Didymos with its European component, the AIM (Asteroid Impact Mission) spacecraft \citep{aimsci}, to perform a hyper-velocity impact on the small moon, called hereafter Didymoon, of the binary with the US component, the DART (Double Asteroid Redirection Test) mission \cite{dartsci}, and to observe the outcome of the impact from AIM. The impact is planned to take place in $2022$ when Didymos is approaching the Earth as close as $\sim 28$ Lunar Distance \citep{obsrvgrp}, so that it can also be observed from ground based observatories.

AIDA will thus be the first large-scale artificial impact on a potentially hazardous binary Near-Earth Asteroid (NEA) with a totally measurable deflection effect (the impact could make a perceptible shift of Didymoon's orbit; see \cite{dartsci} for details), and it provides a unique opportunity to study the cratering process, as well as to understand the ejecta dynamics based on realistic data, including the quantity of the ejecta, their ejection velocity and size distributions, and more importantly, the ejecta's life cycle as a function of these characteristics. 

The ultimate fate of the ejecta released from a crater is a crucial information that has many implications in planetary science. For instance, there is an ongoing debate on the contribution of cratering to the formation of regolith on small asteroid surfaces. As one important outcome of the AIDA mission, the statistical behavior of ejecta produced by the DART impact will be analysed and used to improve the theoretical modeling of the impact process \citep{hols83, schm87, hols93}. Laboratory experiments have been performed for decades, which improved drastically our understanding of the impact process on actual geological materials, but the sizes of sample targets in these experiments, generally of the order of centimeters, are orders of magnitude smaller than the size of actual asteroids \citep{exprmnt1, exprmnt2, exprmnt3}. Scaling laws were developed based on dimensional analysis that indicate how multiple parameters of the impact process relate to each other if the experiment conditions change. \citet{sclform} constructed the basic formulas of mass and velocity distributions of the ejecta, and experiments were then conducted in order to obtain the empirical values of the constant parameters employed in the scaling laws \citep{crtrpara2, crtrpara3}. Several lately developed measuring techniques have been used to determine the unknown constants under different conditions \citep{3dpiv, oblqimpct} and for different physical regimes \citep{regime1, regime2}. However, it is still an open question whether these laboratory results can be successfully extrapolated to a collisional event at planetary or asteroid scale. In particular, it has been established that the strength and porosity of the target surface material relate to the size of the asteroid \citep{grvaggr}. Moreover, \citet{crtrpara1} showed that the ejecta velocities decrease as the target strength decreases, which, as we will demonstrate in this paper, will largely govern the statistical behavior of the ejecta cloud, especially the proportions of escaping, orbiting and re-accreted ejecta. In brief, ejecta with a speed over a critical value will tend to escape directly, unless blocked by the asteroid body, and the rest of the ejecta will be temporarily trapped in cycling orbits around the asteroid(s), which can imply a more complex orbital evolution. 

Dynamical considerations are technically unavoidable in the discussion of the ejecta fate. According to the scene settings of AIDA, the complexity of the mechanical environment near the binary Didymos might be unprecedented in previous space missions (see a detailed discussion in \citet{aidafateI}), especially for the low-speed ejecta that are fated to remain in the vicinity of the binary. Starting from an individual ejecta trajectory, several important factors must be considered, according to the situation. Firstly, the orbital motion is highly influenced by the peculiarities of the binary components in terms of shape, gravity and rotation when the ejecta is approaching the surfaces \citep{dyneros, dynvesta}. Part of these quantities are currently uncertain and will not be known until AIM performs the in-situ characterization of Didymos. Secondly, the trajectories trapped in the neighbourhood of Didymos evolve basically in the context of a Circular Restricted $3$-Body Problem (CR$3$BP) that contains a variety of perturbations that take different forms. Some trajectories initialized closely in phase space may extend across a large space around the binary during a time corresponding to several orbital periods ($P_{orb}=11.92$ hr) of the binary because their Lyapunov times are comparable or even smaller than this time \citep{orbstblya, lyatim3bp}. Thirdly, the perturbation forces acting on the ejecta are due to multiple mechanisms. Collisions between the ejecta are supposed to be dense in the excavating stage, and to become increasingly rare as the dispersion of the curtain increases. \cite{aidafateI} presented a survey of the mechanical environment and found that the solar tide and solar radiation pressure are two major forms that affect the month-long evolution of the ejecta cloud. An informative model was thus created by combining these refined physical processes and known / hypothetical properties of the binary system according to the observations. In fact, the orbital behavior of the ejecta cloud has barely been discussed so far at this sophisticated level. \citet{jimmodel} proposed the excavation flow properties model (EFPM), which was developed to track the fluctuation of the ejecta plume that is extended into a region far from the crater using a semi-empirical initialization. The EFPM was applied to track the ejecta motion from hypothetical impact sites on Didymoon, showing the distribution of escaping ejecta on the celestial sphere and of the accreted ejecta on the surface of the bodies \citep{jimwork}. 

This paper, as the second one of the present series, applies our proposed numerical model to explore the fate of the ejecta in a systematic way, which will allow us to understand the behavior of the ejecta cloud over a wide range of parameters and draw some statistically meaningful interpretations. Section \ref{s:sec2_atlas} describes the initialization of our ejecta grid on the surface of Didymoon, which aims at a broad examination of the fate dependencies on two key factors, the launching site (or projectile impact site) on the grid and the ejection speed. Section \ref{s:sec3_tracing} presents full-scale simulations of the ejecta evolutions from $6$ hypothetical projectile impact locations, considering two material types composing the ejecta to define scaling law material parameters as well as the major orbital perturbations in a $2$-months timescale. Conclusions and a discussion of the results are presented in Section \ref{s:sec4_conc}. 

\section{An Atlas of the Ejecta Fate} \label{s:sec2_atlas}

The analysis of the fate of the debris ejected from an impact should not be only limited to what happens when the ejecta collides with one of the binary components, or escapes from the Sphere Of Influence of the system (abbreviated as SOI; for Didymos, the equivalent radius of the SOI relative to the Sun is $\sim 9.5$ km, and we consider ejecta evolving beyond this distance to escape the system). We also analyze the time sequences preceding these events as a function of launching conditions and ejection speeds. 

After their release from the impact location on the surface of Didymoon, a certain fraction of ejecta with low speeds will fall back directly onto the surface of Didymoon, while the high-speed fraction will tend to escape from the binary system straightly, unless part of it is geometrically blocked by the primary. All these cases are supposed to be terminated within one orbital period ($P_{orb}$) right after the impact. In addition to these simple cases, part of the ejecta with moderate speeds may finally accumulate on the components' surfaces or escape from the system, only after some extended time during which they remain into orbit within or around the binary system. According to simulations, the cycling time can extend up to at least hundreds of periods. This time period spent into orbit inherently gives a measure of the extent of the dynamical association of the orbiting ejecta with the binary system, noting their exposure to a strongly perturbed environment and yet being able to survive. From the analysis of the evolution of the whole ejecta cloud, a relation is found between the launching site on Didymoon as well as the ejection speed of the ejecta and their lifetime and fate, which can be easily represented on a map of Didymoon's surface. In other words, knowing where an ejecta is launched and at which speed, we can statistically determine how long it will evolve and its fate.

Based on this understanding, we classify seven kinds of dynamical fates of ejecta launched from the surface of Didymoon: 

\begin{itemize}
\item[$EE$:] Early Escape -- Escape prior to the first periapsis passage to the binary system. 
\item[$LE$:] Late Escape -- Escapes after the first periapsis passage to the binary, \ie\ the orbit switches from ``cycling'' to ``escaping''. 
\item[$EAP$:] Early Accretion on the Primary -- Impact on the surface of the primary prior to the first periapsis passage to it. 
\item[$ERS$:] Early Re-accretion on the Secondary -- Re-accretion on the surface of Didymoon prior to getting out of the SOI (Didymoon's SOI radius is $\sim 470$ m relative to the primary). 
\item[$LAP$:] Late Accretion on the Primary -- Impact on the surface of the primary after its first periapsis passage to it. 
\item[$LRS$:] Late Re-accretion on the Secondary -- Re-accretion on the surface of Didymoon during a fly-by, after escaping from its SOI. 
\item[$SO$:] Surviving Orbit -- Until a given time limit, the ejecta has neither been intercepted by either of the binary components nor been ejected out of the binary system. 
\end{itemize}

Note that we call accretion an impact of ejecta on the primary and a re-accretion an impact of ejecta on Didymoon from which ejecta originate. Then, we intentionally assume that the ejecta sticks to its position at impact when it impacts either of the components, as if it will form a small crater, which neglects any possible post-impact motion or bouncing. Assuming this is mostly correct, this makes it possible to check the potential contribution of these impacts to surface cratering, \eg\ the distribution of secondary / sesquinary craters related to the AIDA impact \citep{catenae}. The important parameters to be noted are: 1) the escape speed from Didymoon's surface, which is $\sim 8.9$ cm/s assuming a single Didymoon; 2) the escape speed from the binary system at $1.18$ km from the barycenter, which is $\sim 24.4$ cm/s, and 3) the mean orbital speed of Didymoon, which is $\sim 17.3$ cm/s. These values already express some simple facts, \eg\ the ejecta with $v_e \ll 8.9$ cm/s will be re-accreted by Didymoon (ERS), the ejecta with $v_e \gg 50$ cm/s will tend to escape (EE) unless accreted on the primary (EAP), and those with moderate values will have the opportunity to be trapped into temporary orbits in the binary system and end up in various fates. 

The ejection speed $v_e$ is obviously one key parameter that allows a hierarchical  mapping of the ejecta fates. Another key factor is the launching site of the ejecta that largely determines the launching direction. We consider multiple iso-valued ejecta sets, each includes $\sim 100,000$ sampled ejecta particles positioned over Didymoon's globe and assigned a uniform ejection speed. This simulation scheme is designed to illustrate the global distribution of the dynamical fates of sampled particles over a full coverage of Didymoon's surface, and as a function of the ejection speed $v_e$. \citet{aimsci} presented a detailed discussion on the physical properties of the Didymos system within uncertainties, including both the directly measured and the derived parameters. Compatible models were derived for both the primary and Didymoon based on existing parameters and combining the radar and photometric observations \citep{bobsv,pobsv}. We note however that the physical model of Didymoon is still quite hypothetical due to the lack of direct measurements. The heliocentric orbit was then calculated by using the ephemeris at the scheduled deflection time, and the mission scenario was numerically implemented, which enables us to perform the post-impact simulation under a full model (see \cite{aidafateI} for the parameters of Didymos' reference model). This study still relies on this numerical model. Table \ref{t:sampara1} presents the sampling setting parameters of the sampled ejecta particles. For each simulation, assigning the same $v_e$ value for all particles of the set, and tracing the motion of the sampled particles within a specified period, we make an ergodic attempt to search for the dynamical fates of ejecta released in all directions from Didymoon. Then, considering a wide range of $v_e$ from $1$ to $100$ cm/s, the scheme can show systematic patterns of ejecta fates when going through the parameter space represented by the launching site coordinates $\left ( \lambda, \phi \right )$ and ejection speed $v_e$ ($\lambda$ and $\phi$ denote the longitude and latitude of the launching site, respectively; see Appendix A for the definition). 

\begin{center}
Table \ref{t:sampara1}.
\end{center}

$25$ groups of simulations in total were performed to sweep the parameter space and to analyse the dependencies of the ejecta fate on $\left ( \lambda, \phi \right )$ as well as $v_e$. A detailed setup of the system's initial configuration is exposed in Appendix A. A two-month long evolution of the tracer particles was considered. This time is supposed to be long enough for the interest of a space mission, and to understand the contribution of ejecta to surface regolith. In effect, it enables a complete observation, \ie\ most tracer particles may have either impacted one of the two binary components, escaped the system or entered a period of relatively stable motion within the simulated time. A relatively large ejecta size and low material reflection rate (see Table \ref{t:sampara1}) are first chosen on purpose to reduce the effect of solar radiation pressure, so that the results can be viewed as a benchmark for representing the possible outcomes of the ejecta's trajectories perturbed by gravity only. The dependency of the outcome on the solar radiation pressure is the topic of a separate discussion (see Section \ref{s:sec32_srp}). Another assumption of the sampling strategy is that the launching angle with respect to the local horizontal of the surface is fixed to the local normal vector (outwards), not accounting for some non-zero ejection angles like in real cases. This is assumed primarily in order to reduce the dimensions of the parameter space, which restrains the computational cost to an acceptable level. Note that the simulations with low $v_e$ can represent the outcome of possible global processes related to seismic activities induced by a hypervelocity impact or to the secondary / sesquinary impacts. Seismic activities may lift off regolith grains at low speeds in the normal direction due to the low-gravity environment, so that our study can also represent the outcome of this process. 

\subsection{Statistics of ejecta fate } \label{s:sec21_comp}

All the sampled particles with $v_e \le 4.0$ cm/s are found to belong to the class $ERS$ within a very short timescale ($T_f < 1.35$ h), \ie\ ejecta launched with a speed below this limit will re-accrete on Didymoon's surface without ever escaping from its gravitational influence. The final time of our observation $T_f$ is defined as the time when the number of particles in the class $SO$ drops to zero, or when the default simulated time is reached ($60$ days for this study). On the other hand, none of the sampled particles with $v_e \ge 42.0$ cm/s is found to eventually re-accrete on Didymoon (neither in class $ERS$ nor in class $LRS$). More precisely, when $v_e = 42.0$, $97.57\%$ of the particles fall in the class $EE$, and only $2.43\%$ of the particles that are launched towards the primary (around $\lambda = 164.8^\circ$ W, $\phi = 0.2^\circ$ S; see Fig. \ref{f:fateatlas}) will accrete on the primary (class $EAP$). We calculated the $3\sigma$ limits of the cleanup procedure of orbiting ejecta and the relative proportions of dynamical fates: given an ejection speed, all sampled particles are defined in class $SO$ at the moment of release, and the proportion of $SO$ particles shows a monotonous decline due to the increasing fractions accreted on either of the binary components or leaking out of the SOI. Then a $3\sigma$ rule is applied to mark the decreasing curves of the $SO$ particles, \ie\ the time limits when $68\%$, $95\%$ and $99.7\%$ of the $SO$ ejecta had a final fate (denoted as $T_{1\sigma}$, $T_{2\sigma}$, $T_{3\sigma}$, respectively) were recorded and the proportions of the fates, including those ``accreted'' ($EAP$/$ERS$/$LAP$/$LRS$) or ``escaped'' ($EE$/$LE$), were calculated at times statistically ruled as $T_{1\sigma}$, $T_{2\sigma}$ and $T_{3\sigma}$. Table \ref{t:glbres} presents the ruled values of these percentages for all the $25$ simulations, together with the proportions of the fates at the final time. 

\begin{center}
Table \ref{t:glbres}.
\end{center}

The ``$68$-$95$-$99.7$'' rule serves as a measurement of the diffusion process of the ejecta orbits, and further provides a rough probability estimate of the various possible ejecta fates. We then find that none of the ejected particles with ejecta speeds equal to $4.0$, $4.5$ and equal to or greater than $42$ cm/s are still orbiting in or about the system at the termination time (see Table \ref{t:glbres}). Other ejecta with ejection speeds in the range $5.0$--$40.0$ cm/s have a more varied distribution of fates. Out of those ranges, the fate of the particles is more systematic. In particular, when the ejection speed $v_e$ is below $4.0$ cm/s and approaching $0.0$ cm/s, the trajectories of the particles relative to Didymoon will tend to take the form of a small loop that goes up and back to Didymoon; when $v_e$ is above $42.0$ cm/s and approaching to the maximum (the upper limit of the ejecta speed can reach the magnitude of $10^2$ m/s according to experimental data \citep{dartsci}, which highly depends on surface material properties), the trajectories of sampled particles will tend to be radiating in launching directions, causing a rapid clearance of ejecta from the vicinity of the binary system, except for a small fraction of the trajectories (the particles launched against the primary) that will end up on the primary's surface because of the occultation. 

The rapid cleanup of $SO$ particles occurs for ejection speeds in the range $4.0 - 5.0$ cm/s and above $38.0$ cm/s, corresponding to an average $2\sigma$ time below $2$ days within which the proportion of orbiting ejecta drops by at least $95\%$. From $6.0$ cm/s to $10.0$ cm/s, the average cleanup time (both $T_{1\sigma}$ and $T_{2\sigma}$) shows a sharp increase, \ie\ particles with such ejection speeds have higher probability to survive in the environment of the binary system for a longer time, \ie\ up to tens of days. The residual proportion of $SO$ particles peaks in ejection speeds between $12.0$ cm/s and $18.0$ cm/s. In this range of ejection speeds, over $5.0\%$ of the sampled particles will remain orbiting about the system at the final time. For ejection speeds from $20.0$ cm/s to $36.0$ cm/s, the average cleanup time ($T_{1\sigma}$ and $T_{2\sigma}$) exhibits a uniform decrease, which statistically means that the particles with higher speeds show more instability in their  temporarily trapped trajectories in Didymos' vicinity.

We checked the relative proportions of fates as a function of the ejection speed $v_e$. At all $4$ test instants $T_{1\sigma}$, $T_{2\sigma}$, $T_{3\sigma}$ and $T_f$, the proportion of escaping particles ($EE$/$LE$) show a monotonous increase with the increase of $v_e$, which is consistent with intuitive expectations. Meanwhile, the number of particles accreted on the primary ($EAP$/$LAP$) behaves up and down at the $4$ instants, apart from the $v_e$ values where the $2\sigma$/$3\sigma$ limits are not applicable (see Table \ref{t:glbres}). The proportion of fates ending on the primary also peaks between $12.0$ cm/s and $18.0$ cm/s, which exactly matches the variation of the number of residual $SO$ particles at $T_f$. There is a common mechanism behind this correlation: the high residual percentage presents a survival advantage in statistics, \ie\ trajectories of sampled particles with ejection speeds from $12.0$ cm/s to $18.0$ cm/s spend longer time in average orbiting close to the primary than particles with other ejection speeds; because orbits around the primary are perturbed by an irregular potential, orbiting particles increase their chances to accrete on the primary's surface. The proportion of ejecta re-accreted on the secondary ($ERS$/$LRS$) shows an overall trend of decrease with the increase of $v_e$, which is primarily resulting from the increasingly lower proportion of orbiting ejecta with ejection speed. However, there is an exception, \ie\ a small peak is noticed in the proportion of ejecta re-accreted on Didymoon for ejection speeds between $30.0$ cm/s and $35.0$ cm/s (at both $T_{1\sigma}$ and $T_f$), which appears as a small rise and resumes to decline quickly. A more complex mechanism is involved for this anomaly, which involves a resonance with Didymoon's orbital period that appears to play a crucial role (see Section \ref{s:sec22_reso}). 

\subsection{Dynamical properties of accreting/re-accreting ejecta} \label{s:sec22_reso}

The $7$ types of fates of the sampled particles are mapped against the launching sites over the surface of Didymoon (Fig. \ref{f:fateatlas}), and the patterns in terms of different ejection speed levels reveal rather sophisticated structures. The different times at which the map are shown after ejection, at early days ($3 - 6$ days), have been chosen so that the interesting patterns show up clearly before being mixed at later times. The snapshots of Fig. \ref{f:fateatlas} illustrate $6$ representative cases from our numerical search over ejection speeds in the range $4.0 - 42.0$ cm/s. In  the following, we describe these $6$ cases.

Figure \ref{f:fateatlas}(a) shows that $ERS$ ejecta only exist for ejection speeds below $10.0$ cm/s, and the area on Didymoon within which ejecta become $ERS$ shrinks with increasing $v_e$. Ejecta launched at low ejection speed that do not become $ERS$ are first launched from Didymoon's surface locations around $(0^\circ W/E, 0^\circ S/N)$ and $(180^\circ W/E, 0^\circ S/N)$, \ie\ the ejecta from both the proximal and distal areas to the primary can escape from the gravitational influence of Didymoon itself at relatively low $v_e$ and reside in class $EAP$/$LAP$/$LRS$/$LE$/$SO$. It is notable that $EE$ is missing in this case because at low ejection speeds, no ejecta can achieve the escape speed of the whole binary system. 

Figure \ref{f:fateatlas}(b) shows that the area of Didymoon's surface from which $EE$ ejecta are launched exists only for ejection speeds above $8.0$ cm/s and is located firstly around $(50^\circ E, 0^\circ S/N)$ and expands with increasing $v_e$, as $v_e$ gets closer to or greater than the system escape speed. Similarly, on the other side of Didymoon, around $(130^\circ W, 0^\circ S/N)$, ejecta launched from a wide area become $EAP$, which is due to the fact that the ejecta gain lower speeds with respect to the primary that put them on quasi-elliptic trajectories with pariapsis within the primary. Consequently they directly impact the primary's surface. Beyond these areas, all the ejecta will fall in class $LAP$/$LRS$/$LE$/$SO$. 

Figure \ref{f:fateatlas}(c) shows that as $v_e$ is going up from $10.0$ cm/s to $18.0$ cm/s, the area of $EE$ continues to spread over the whole eastern hemisphere. This means that ejecta launched from the hemisphere on the leading direction of Didymoon and with ejection speeds over $20.0$ cm/s will tend to escape straightly. Meanwhile, the launching area occupied by $EAP$ ejecta is also expanding and reaches its maximum extent for ejection speeds between $18.0$ cm/s and $22.0$ cm/s, \ie\ it covers the surface coordinates $[30^\circ W, 180^\circ W] \times [35^\circ S, 35^\circ N]$. Ejecta from the other areas belong to classes $LAP$/$LRS$/$LE$/$SO$. 

Figure \ref{f:fateatlas}(d) shows that as $v_e$ increases again up to $30.0$ cm/s, the launching area of ejecta in the class $EE$ continues to spread to the western hemisphere. In addition, the launching area of the other $5$ types shrinks around the surface coordinates $(90^\circ W, 0^\circ S/N)$. Note that the launching area of $EAP$ is also shrinking at the same time, and as $v_e$ increases from $16.0$ cm/s to $30.0$ cm/s, a launching area of $LAP$ ejecta appears near the equator between $[0^\circ W, 60^\circ W]$, which corresponds to the trajectories missing the primary during the first approach and then accrete on it gradually under the action of Didymoon's gravity. 

Figure \ref{f:fateatlas}(e) shows that as $v_e$ increases from $30.0$ cm/s to $36.0$ cm/s, the launching area of $EAP$ rapidly shrinks to a region around $(165^\circ W, 0^\circ S/N)$, where the ejecta are launched towards the primary. Then, the ejecta that are launched from the nearby areas will be transferred into temporary orbits around the primary, which gives them more chances to cross Didymoon's trajectory and therefore re-accrete on the moon. The sharp decrease of $EAP$ launching area with ejection speeds within $30.0 - 36.0$ cm/s reveals the cause of the single peak of the curve of ejecta re-accreted on Didymoon for these particular ejection speeds: at  lower ejection speeds, all ejecta collide directly with the primary, then as the ejection speed increases within the indicated range, the ejecta tend to miss the primary and rather pass closely by the object and eventually re-accrete on Didymoon. This explains the small peak in the number of re-accreted ejecta for these ejection speeds.

Fig. \ref{f:fateatlas}(f) shows that for $v_e$ equal to $38.0$ cm/s, the launching area of $EAP$ is well separated from the launching area of ejecta that will temporarily be in orbit within the binary system or re-accrete on Didymoon. The former area resides around $(165^\circ W, 0^\circ S/N)$, while the latter area is around $(90^\circ W, 0^\circ S/N)$. In this latter area, only a small fraction of the ejecta is transferred onto temporary orbits, the rest re-accretes on Didymoon. Increasing $v_e$ would actually shrink this latter area, without changing much the one of $EAP$. Note that ejecta launched at relatively high speed accrete very rapidly on the primary.

\begin{center}
Figure \ref{f:fateatlas}.
\end{center}

The patterns shown on the snapshots describe well the complex dynamics involved in the ejection process from Didymoon's surface. The collisional conditions between the ejecta trajectories and Didymoon's orbit are key to understand these patterns. Unfortunately, these conditions are not as obvious as in the collisional analysis of Keplerian orbits \citep{disorb}, because the perturbations (third-body attraction, non-spherical gravitational potentials, tidal forces, solar radiation pressure, \etc) introduce noises in the dynamics, which result in frequent unpredictable changes of the orbital characteristics during the observation time. Nevertheless, two major mechanisms are still noticed to be working broadly: orbital resonances and non-resonant collisions. 

Resonances between the orbital periods of the ejecta and Didymoon prove to be a plausible long-term mechanism for the re-accretion process on the secondary. When the orbital period of Didymoon $P_{orb}$ and that of the ejecta around the primary are commensurate, a periodic gravitational influence of Didymoon on the ejected particles occur. The result is an increase of the perturbation of Didymoon on the ejecta, which can either lead to a collision with Didymoon or drastically modify the orbits of the ejecta. Collisions first occur for ejecta having an orbital period related with $P_{orb}$ by a ratio of two small integers, \ie\ after being ejected from Didymoon's surface, this kind of ejecta will rendezvous with Didymoon after a few orbits only (the number depends on the resonant ratio) and get re-accreted on its surface. The launching sites of sampled ejecta governed by resonances follow regular patterns on the surface map of Didymoon, hereafter named resonant bands, which are within the launching area occupied by the $SO$ type of ejecta because only the temporarily orbiting ejecta respond to this mechanism. The shapes of the resonant bands can be approximated through a simplified model: omitting the gravity from the secondary, the launching sites subject to $m/n$ resonance distribute on the ring defined by Eq. (\ref{e:resoeqn}) ($m/n$ means the ratio of the ejecta's period over $P_{orb}$). 

\begin{equation} 
\label{e:resoeqn}
1 - 2 \hat{v}_e \textup{cos} \phi \ \textup{sin} \lambda -  \hat{v}_e^2 \approx \left ( \frac{n}{m} \right )^{2/3}, 
\end{equation}

in which $\hat{v}_e$ indicates the ejection speed normalized by the average orbital speed of Didymoon, in the radial direction defined by $\lambda$ and $\phi$. This approximation works for relatively large values of $v_e$ that enable ejecta to be ejected on detached orbits from Didymoon. For example, in Fig. \ref{f:fateatlas}, the shapes of resonant bands are systematically observed when $v_e \ge 10$ cm/s. We then examined the time variation of the $6$ snapshots over $3$ days of evolution ($\sim 6\ P_{orb}$), and found that resonances occur for a large range of orbital period ratios. In the case of Fig. \ref{f:fateatlas}(b), we find the resonant bands with $m/n$ equal to $1/1$, $1/2$, $2/3$, $3/4$, $3/5$, $4/3$, $4/5$, $5/4$, $5/6$, $6/5$ are located in the western hemisphere, and those with $2/1$, $3/1$, $3/2$, $4/1$, $5/1$, $5/2$, $5/3$, $6/1$ are in the eastern hemisphere; likewise in the case of Fig. \ref{f:fateatlas}(c), the resonant bands with $m/n$ equal to $1/1$, $2/1$, $2/3$, $3/1$, $3/2$, $3/4$, $4/3$, $4/5$, $5/2$, $5/3$, $5/4$, $5/6$, $6/5$ are located in the western hemisphere, and those with $4/1$, $5/1$, $6/1$ are in the eastern hemisphere (the rings are shifting as the $EE$ area expands); in the case of Fig. \ref{f:fateatlas}(d), all the resonant bands are found in the western hemisphere with $m/n$ equal to $1/1$, $2/1$, $2/3$, $3/1$, $3/2$, $3/4$, $4/3$, $4/5$, $5/2$, $5/3$, $5/4$, $5/6$ and $6/5$; in the case of Fig. \ref{f:fateatlas}(e), all the resonant bands are found in the western hemisphere with $m/n$ equal to $1/1$, $2/1$, $3/1$, $3/2$, $4/1$, $4/3$, $5/1$, $5/2$, $5/3$, $5/4$, $5/6$, $6/1$ and $6/5$; finally, in the case of Fig. \ref{f:fateatlas}(f), only the resonant bands with $m/n$ equal to $2/1$, $3/1$, $4/1$, $5/1$, $6/1$ are found inside a small launching area in the western hemisphere. Note that we just performed this analysis over a time corresponding to $6\ P_{orb}$ and it may be that other resonances (corresponding to different ratios) occur at later time. The reason of fixing this time limit is because within this time, the sampled ejecta that are trapped in a resonance with the identified ratios are packed regularly and largely follow the shapes and orders predicted by Eq.~(\ref{e:resoeqn}). This signature will be lost at later time, because of the accumulation of perturbations that will prevent us to check the resonance bands in this way, even if they are still occurring. In fact, for all integer ratios of orbital periods, the effects of orbital resonances can take place over a long time and lead to pronounced quasi-periodic re-accretion peaks (collisions). For the cases discussed above, the duration of this sustained effect is from $30$ days up to $60$ days. 

Although the mechanism can still be active after tens of $P_{orb}$, the launching sites of the ejecta that re-accrete and form the peak at late times are not organized as regularly on Didymoon map as those that do so within the first few $P_{orb}$. They rather get blurred and dispersed across the whole launching area of $SO$ ejecta (marked in green in Fig. \ref{f:fateatlas}). This is understandable because chaos in CR$3$BP essentially grows over the observation time (even if the exact relationship between chaos and Lyapunov time is not straight; see \citet{winwin}). Specifically in our cases, the diffusion of resonant bands is correlated with the close passing by of ejecta affected by the early resonances (with simple ratios as stated above), \ie\ ejecta belonging to the early bands re-accrete within a few periods, while the ejecta launched with very close initial conditions but outside the resonance bands survive by passing by Didymoon closely and gain a rapid orbital change due to the strong gravitational interaction with the moon. The change in ejecta trajectories through this swing-by effect can be quite drastic and its amplitude highly depends on the exact configuration of the encounter between Didymoon and the ejecta and of the binary components at the time of the encounter. A sharp increment of the orbital energy is usually due to the interplay with Didymoon, which leads to two distinct tendencies: 1) a positive increment, which can cause an escape of the ejecta from the binary system if its speed with respect to the binary system posterior to the encounter exceeds a critical value; 2) a negative increment, which both decreases the orbital period but also increases the eccentricity, and which causes the ejecta to fall directly onto the surface of the primary if the negative increment is high enough . These considerations are perfectly supported by the results of our simulations. A big fraction of both $LAP$ and $LE$ ejecta are noticed to be launched from areas that keep close to the early resonant bands (Fig. \ref{f:fateatlas} (b)--(e)). For the other ejecta that gain insufficient energy increments to make them belong to the $LAP$ or $LE$ classes, the temporary orbiting status remain, and like a cycle, the next significant orbital change will occur when the ejecta are approaching Didymoon again. In particular, there exists a small fraction of the ejecta passing by the secondary, which gain basically no energy increment, \ie\ their orbit is changed to another one that remains close to and crosses Didymoon's orbit. For the ejecta trapped in a mean motion resonance with Didymoon, this leads to a resonant return to its vicinity with exactly equal time intervals. That is the reason why we observe in Fig. \ref{f:fateatlas} that some early resonant bands ($n < 3$) become reinforced and grow thicker right after another $n \times P_{orb}$. 

Non-resonant collisions are another common outcome that can produce even more diversified patterns on Didymoon maps, because the collision conditions in this case imply more solutions that are distributed over a wider parameter space, and therefore many launching areas. In a simplified analysis, the distance between arbitrary unperturbed orbits was well formulated by \citet{disorb}, and accordingly the critical points of the distance function can be solved by finding the real roots of a trigonometric polynomial. So a fraction of the ejecta may encounter with Didymoon in any orbital phase corresponding to non-integer ratios of the periods,which are also most likely inclined orbits from the equatorial plane. Differing from the resonant cases, the non-resonant collisions are not recurrent, and most patterns of ejecta in the $LRS$ class created by the non-resonant collisions appear as block shapes on Didymoon map (Fig. \ref{f:fateatlas}), and some of these blocks are asymmetric about the equator (Fig. \ref{f:fateatlas} (b)). On the other hand, like in orbital resonances, the swing-by effect also works for the non-resonant case, as we also notice concentrations of $LAP$ ejecta  (Fig. \ref{f:fateatlas} (a)--(e)) and ejecta from the $LE$ class (Fig. \ref{f:fateatlas} (a)--(b)) adjacent to the $LRS$ blocks. Moreover, we note that for the high $v_e$ range ($> 36$ cm/s), the ejecta of classes $LAP$ and $LE$ almost vanish, namely the orbital energy remains constant unless the ejecta re-accrete on Didymoon, which supports the theoretical anticipation that retrograde orbits have a higher stability and therefore a larger advantage to survive. 

We also analyzed the impact speeds of the accreted ejecta on the primary and re-accreted ones on Didymoon for all sampled ejecta in class $EAP$, $ERS$, $LAP$ and $LRS$. The impact speeds are derived via interpolation at the instance when collisions with the components' surfaces are detected, and are represented in the local frames (see Appendix A). Figure \ref{f:reimpact} shows the total range of impact speeds on the surfaces of both the primary and the secondary found in our simulations. The boundary curves mark the upper and lower limits of the whole range, respectively, which statistically gives a measure of the extent of the energies of impacts by ejected debris as a function of their ejection speed. 

\begin{center}
Figure \ref{f:reimpact}.
\end{center}

The average impact speed shows a monotonous increase with rising ejection speeds $v_e$ between $4.0$ cm/s and $55.0$ cm/s, which is similar for both the accreted particles on the primary ($EAP$/$LAP$) and the re-accreted ones on the secondary ($ERS$/$LRS$). We also find that the range of impact speeds on the secondary grows and then shrinks over this domain of ejection speeds, and the limit curves converge to the line corresponding to equal ejection and impact speeds (Fig. \ref{f:reimpact}) as $v_e$ is approaching $4.0$ or $40.0$ cm/s. The range width peaks around $v_e = 12.0$ cm/s, which is the value of launching speeds that causes $LRS$ ejecta to follow a great diversity of dynamics (Fig. \ref{f:fateatlas} (b)) and to have a relatively long survival time (Table \ref{t:glbres}). In comparison, the range of impact speeds on the primary proves to be much wider ($> 24$ cm/s), and its width shows a positive correlation with $v_e$. This wide range is mainly the result of the high spin rate of the primary, which produces a $\sim 30$ cm/s speed at the edge of the equator. \citet{jimwork} indicated that a high spin rate results in a size-sorting of the ejecta of different sizes. Here we find that ejecta launched from close launching sites may end up on the surface with very different impact speeds just because the impact points are apart. 

The results can be used to estimate potential geological changes due to the impacts of ejecta. The low speed impacts on the primary are limited to low latitudes, \eg\ impacts below $15$ cm/s are restricted to $[40^\circ S, 40^\circ N]$, and impacts below $20$ cm/s are restricted to $[60^\circ S, 60^\circ N]$. The polar regions are only affected by impacts around $40$ cm/s. Then above $70$ cm/s, the impact locations converge again to $[60^\circ S, 60^\circ N]$. As for Didymoon, Fig. \ref{f:reimpact} shows that the maximum impact speed is about $40$ cm/s, and as the impact speed increases up to that value, the possible impact locations are highly weighted towards the eastern hemisphere, \ie\ high-speed impacts are caused by ejecta impacting from the leading direction of Didymoon. 

\subsection{Analysis of the survival orbits} \label{s:sec23_soana}

To evaluate the orbital distribution of the $SO$ particles at the termination time ($60$ days), we derived the instantaneous orbital elements of these particles using their cartesian positions and velocities calculated with respect to the translational frame originating at the centre of mass of the primary. The number of survived ejecta at $T_f$ for each ejection speed are indicated in Table \ref{t:glbres}. As listed in this table, the $SO$ type exists only for $5.0$ cm/s $\le v_e \le 40.0$ cm/s and the number shows a ``rise and fall'' change. Since the trajectories at $T_f$ have entered a steady stage of evolution, \ie\ the shapes of approximated Keplerian orbits exert slow variation and the instantaneous elements actually reflect the distribution of the orbits of survived ejecta. Figure \ref{f:orbdist} illustrates the distribution of the orbital elements at $T_f$, projected into two parameter planes, respectively. The semi-major axis vs. eccentricity distribution shows that the eccentricities of survived orbits cover a wide range, and the semi-major axis is relatively concentrated within a range up to just a few times that of Didymoon. The longitude of ascending node vs. inclination shows a remarkable dependency of the survived orbits on the ejection speed. More precisely, ejecta with the same $v_e$ are concentrated in two clumps in this diagram located in two distinct areas, whose centers have their longitudes of ascending node separated by $\sim 180^\circ$ and inclinations separated by $\sim 10^\circ$. 

\begin{center}
Figure \ref{f:orbdist}.
\end{center}

Figure \ref{f:orbdist} reveals several features of the orbits of surviving ejecta. 1) Particles with lower $v_e$ have a greater semi-major axis, and when the semi-major axis exceeds $2.5$ km, the orbital eccentricity is also high ($>0.4$). Moreover, we find that when $v_e$ exceeds $36$ cm/s, all the survived orbits have an eccentricity greater than $0.4$. 2) The orbits of surviving ejecta with the same $v_e$ value are found to be concentrated in narrow ranges of inclination, and the averaged level declines as the ejection speed increases. For the nearly equatorial orbits (see clumps a--c and i in the right plot of Fig. \ref{f:orbdist}), both prograde and retrograde, the survived orbits cover the full equatorial area, and leave a conic vacuum of ejecta in the polar areas. For the orbits with large inclinations (clumps d--h), the double-clumps turn to be more densely occupied along the axes. These pairs of discs, whose relative orientation is shifted by $\sim 10^\circ$, cover only a certain fraction of the polar areas, and leave a vacuum of ejecta between them (see the blank region on the right plot of Fig. \ref{f:orbdist}). This result being based on a quasi-ergodic search, we can conclude that the double disc is actually a universal structure in the ejecta cloud dynamics, which exhibits good stability under the dynamical environment of the binary system. The vacuum region could be a valuable reference for the mission design of a spacecraft observing an impact on Didymoon and planed to return in the close vicinity of the system after $T_f$. 3) The results confirm the survival advantage of polar orbits and retrograde orbits. We then analyzed the distribution of survived orbits as a function of inclination. We find that $59\%$ of survived orbits fall in the inclination range $[120^\circ, 180^\circ]$ (\ie\ retrograde, nearly equatorial orbits) and consist mostly of particles with low $v_e$ values. Then, $36\%$ of survived orbits fall in the inclination range $[60^\circ, 120^\circ]$ (\ie\ polar orbits). Only $5\%$ of survived orbits have an inclination below $60^\circ$ (\ie\ prograde nearly equatorial orbits) and are mostly composed of particles with large $v_e$ values that enter highly eccentric orbits after being ejected. 

\section{Full-Scale Simulation of Ejecta Cloud Evolution} \label{s:sec3_tracing}

\subsection{Initialization and scene settings} \label{s:sec32_iniscn}

A fascinating effort for the study of ejecta cloud is to track their motion throughout the whole process, \ie\ from the initial stage of cratering to the final stage of accretion or escape. This will provide crucial information that may help to clarify the role of impact ejecta in the formation of regolith on planetary and small body surfaces. It also guides the strategy of observations of an observing spacecraft of an impact event, as the AIDA project proposes to do. The predictions rely on our knowledge of the ejection mechanics of a kinetic impact at the appropriate scale, which to date has not been verified at asteroid scales. Theoretical models, usually defined as scaling laws, have established dimensionless relationships between the impact conditions, target properties and the regularized outcomes. As in \citet{aidafateI}, to define the ejecta initial conditions, we adopt the scaling laws developed by \citet{crtrpara3}. Based on their descriptions of the analytical forms of the ejection mass and speed, \citet{dartsci} presented the scaling laws in the context of the AIDA mission. In this section, we employ the previously developed two-stage numerical methodology \citep{aidafateI} to track the evolution of ejected particles initialized using the data exported by the scaling laws. To be consisent with the DART mission, the projectile is modeled as a $500$ kg sphere of radius $0.5$ m, and the normal incident speed is $6$ km/s. \citet{dartsci} estimated the appropriate range of the target material strengths based on the $8$ experimental media used in \citet{crtrpara3}, and assumed that the weakly cemented basalt (WCB, low porosity, moderately high strength) may be a good representative of Didymos primary, while perlite/sand mixture and sand/fly ash (PS \& SFA, both high porosity and low strength) may be good analogues for Didymoon. These two kinds of material, WCB and SFA, are chosen for our simulations. Since the actual properties are unknown to date, this choice may appear deliberate but we assume that it is representative of the wide possible range of surface/subsurface properties of Didymoon. For each case, $6$ hypothetical projectile impact sites on Didymoon's surface are considered to cover all vertices of the ellipsoidal shape. More precisely, $4$ sites locate on the equator at longitude, latitude of $(0^\circ, 0^\circ)$, $(90^\circ W, 0^\circ)$, $(180^\circ, 0^\circ)$, $(90^\circ E, 0^\circ)$, and $2$ sites locate at the poles $(0^\circ, 90^\circ S)$, $(0^\circ, 90^\circ N)$. Polar impact sites are regarded as typical cases of high-inclination impacts, although DART spacecraft's trajectory is now fixed and not planed to reach the pole. However, we consider all possible impact sites, as a general investigation that can also apply to natural impacts. Table \ref{t:materials} lists the scaling parameters of WCB and SFA described by \cite{crtrpara3}. 

\begin{center}
Table \ref{t:materials}.
\end{center}

The crater radius $R$ can be calculated by Eq. (\ref{e:sclR}), which evaluates the eventual crater radius in the ``strength regime'': 

\begin{equation} 
\label{e:sclR}
R \left( \frac{\rho}{m} \right)^{1/3} = H_2 \left( \frac{\rho}{\delta} \right)^{(1 - 3 \nu) / 3} \left( \frac{Y}{\rho U^2} \right)^{-\mu/2}, 
\end{equation}

in which $a$, $m$, $\delta$, $U$ are the radius, mass, bulk density and impact speed of the projectile, respectively, while $Y$ and $\rho$ are the target's strength and density, respectively, $\mu$ and $\nu$ are material parameters and $H_2$ is a constant. The distribution of the ejection mass $M$ and ejection speed $v$ as functions of the radial distance $x$ from the center of the crater are expressed by Eq. (\ref{e:sclM}) and Eq. (\ref{e:sclV}): 

\begin{equation}
\label{e:sclM}
M \left ( < x \right ) = m \frac{3 k}{4 \pi} \frac{\rho}{\delta} \left [ \left ( \frac{x}{a} \right )^3 - n_1^3 \right ],\ n_1 a \leq x \leq n_2 R; 
\end{equation}

\begin{equation}
\label{e:sclV}
v = U C_1 \left [ \frac{x}{a} \left ( \frac{\rho}{\delta} \right )^\nu \right ]^{- \frac{1}{\mu}} \left ( 1- \frac{x}{n_2 R} \right )^p,\ n_1 a \leq x \leq n_2 R. 
\end{equation}

where $C_1$, $n_1$, $n_2$ and $k$ are each a constant.

\citet{itokwreg} measured the cumulative size-distribution of the granular material on the surface of Itokawa based on the high-resolution images returned by HAYABUSA spacecraft, and found that a power law matches a wide range of sizes, \ie\ from the fine-size pebbles ($<0.05$ m) to large boulders ($>5$ m). The log-log slope is about $-2.8$, which gives an estimate of the amount of cumulative number $N$ of regolith grains, as expressed in Eq. (\ref{e:pwlw}). 

\begin{equation}
\label{e:pwlw}
N (>d) = N_r d^{-2.8},\ d_l \le d \le d_u, 
\end{equation}

where $N_r$ is the reference number of the measured grains, while $d_l$ and $d_u$ determine the lower and upper limits of the full size range, respectively. Equations (\ref{e:sclR})--(\ref{e:pwlw}) define a unique continuous distribution of the scaled ejected material, and provide a mathematical description of the initial phase of the ejecta cloud. The numerical method employed here requires to discretize the ``continuum'' ejecta into the same quantity of individual particles to be used in simulations. In the first paper of the current series \citep{aidafateI}, a piecewise scheme was applied to implement the discretisation, \ie\ the range of the particle size and radial distance were divided into several intervals, and each interval was filled up with particles randomly generated within the specified particle size and mass fraction (see \citet{aidafateI}). An obvious disadvantage of this scheme is the resulting layered ejecta cloud caused by the piecewise structure, which fails to capture the patterns of the real ejecta cloud. In this study, an advanced scheme is developed to realise the discretisation based on a random number generator and the inverse transformation of the distribution function. The scheme was confirmed to be division-free over the full parameter range, and the generated ejecta particles prove to fit Eqs (\ref{e:sclR})--(\ref{e:pwlw}) well (see Appendix B for a detailed description). 

\begin{center}
Table \ref{t:discret}.
\end{center}

Table \ref{t:discret} presents a summary of the discretisation results for both material cases. For each case, the full size range is divided into subranges of different magnitudes that contain different amounts of ejecta particles, which represents well the size distribution of the ejecta cloud. The number of ejecta to follow in simulations can be reduced, noting that the majority of the particles have speeds much greater than the escape speed of the system in both cases. Therefore they obviously belong to class $EE$. Consequently, we limit the simulated number of particles of these classes to a subset. We define a critical speed $v_{cr} = 60$ cm/s based on the results in Section \ref{s:sec2_atlas}. Table \ref{t:sampara2} sets the sampling parameters from the size subranges and for ejection speeds in ranges over and below $v_{cr}$. We consider a greater sample size for the low-speed fraction ($v<v_{cr}$) because the ejecta trapped in the binary system bear more profound implications to the global dynamics of the whole ejecta cloud. We regard the amount of the sampled particles as statistically meaningful although the fraction it represents from the total is relatively small.   

\begin{center}
Table \ref{t:sampara2}.
\end{center}

Simulations have been performed over $60$ days of simulated time. For the evaluation of the solar radiation pressure, Didymos being an S-type whose meteorite analogue is  ordinary chondrites, the albedo of individual particles is assumed to be the one measured for ordinary chondrites, $0.50$, \citep{albedos}. A typical launching angle of $45^\circ$ is assigned to all the sampled particles. 

\subsection{Results} \label{s:sec31_res}

Table \ref{t:fullres} presents the final states of the $12$ full-scale simulations. The momentum enhancement factor $\beta$ \citep{hypvel} can be estimated using the ejection speeds of the sampled particles at the early excavating stage, which, for WCB, gives $\beta \approx 1.11$ and for SFA,  gives $\beta \approx 1.33$. The equivalent impulse and impulse moment are calculated correspondingly and applied to the full binary model, which results in different mutual orbital modifications for the $12$ cases. Table \ref{t:discret} shows that the absolute majority of the ejecta reach much greater speed than the critical value $v_{cr}$ and therefore belong to the $EE$ class. Consequently, the subsequent changes of the secondary's orbit caused by the gradually re-accreted ejecta are neglected (unless the impact triggers a global deformation of either of the binary components, which cannot be discarded but is still a subject of study). 

\begin{center}
Table \ref{t:fullres}.
\end{center}

In all the $12$ cases, the fraction of ejecta accreted or re-accreted on the binary components is confined to low levels. For WCB, as derived from Table \ref{t:fullres}, the proportion of M.A.P. ranges from $2.0 \times 10^{-6}$ to $1.4 \times 10^{-4}$, and that of M.A.S. ranges from $2.6 \times 10^{-7}$ to $5.7 \times 10^{-6}$; for SFA, the proportion of M.A.P. is from $2.6 \times 10^{-5}$ to $2.4 \times 10^{-3}$, and that of M.A.S. is from $7.3 \times 10^{-6}$ to $1.7 \times 10^{-5}$. One reason for these low proportions is that the high-speed ejecta curtain that includes the majority of the ejecta mass is not intercepted by the primary (in all the $12$ cases). Another important reason is that the distribution of ejection speeds derived from the scaling law (for both WCB and SFA) are highly weighted towards large values and therefore, the low-speed ejecta curtain only contains a small fraction of the total mass. However, we need to express some words of caution regarding low-speed ejecta. As shown in \citet{crtrpara3}, scaling laws break at low ejection speeds. Moreover, low-speed ejecta are very difficult to measure experimentally, noting that in the case of Didymoon, the escape speed is tiny (8.9 cm/s); therefore ejecta with ejection speeds as small as a few cm/s can actually contribute to the cloud and are also the ones that can best contribute to the fraction of ejecta that accrete/re-accrete. In this paper, we assume a given proportion of such ejecta, so that we have a reference case.

Taking a detailed look at $6$ possible impact sites of the projectile, we find that in the cases of $(90^\circ W, 0^\circ)$ and $(180^\circ, 0^\circ)$, a greater number of ejecta accrete/re-accrete on the binary components and their ejection speeds $v_e$ cover a wide range of values: $4.5$ cm/s to $59$ cm/s for particles landing on the primary, and $1.5$ cm/s to $58$ cm/s for particles landing on the secondary. In the cases of $(90^\circ E, 0^\circ)$ and $(0^\circ, 0^\circ)$, a lower number of ejecta is found to accrete/re-accrete and their ejection speeds cover a narrower range of values: $5.0$ cm/s to $30$ cm/s for particles landing on the primary, and $1.5$ cm/s to $26$ cm/s for particles landing on the secondary. For the poles $(0^\circ, 90^\circ S)$ and $(0^\circ, 90^\circ N)$, we find an intermediate regime in terms of number of ejecta and range of ejection speeds: the ejection speeds of particles landing on the primary range from $5.7$ cm/s to $38$ cm/s, and that of particles landing on the secondary range from $1.5$ cm/s to $38$ cm/s. This finding is consistent with the results of the grid search (Fig. \ref{f:fateatlas}), which showed that ejecta from these locations are constrained to different $v_e$ limits to be accreted/re-accreted by the binary components, in the same order as listed above. In particular, we see that for both materials, $(90^\circ W, 0^\circ)$ is the best condition to produce regolith material from impact ejecta. 

\begin{center}
Figure \ref{f:wcbacc}.
\end{center}

The two mechanisms discussed in Section \ref{s:sec22_reso} are still found to be active in these full-scale simulations, in spite of the perturbations from the solar radiation pressure that were not accounted for in this Section and the fact that the launching site of sampled particles only covers a small fraction of Didymoon's surface. Defining the accreting/re-accreting rate as the ejecta mass accreted/re-accreted during a short time interval, the recurrent accretion/re-accretion peaks may last for tens of $P_{orb}$. Figure \ref{f:wcbacc} presents an example of the re-accreting rate on the secondary in the case of material WCB. The curves indicate the time variation of re-accreting rate for the $6$ hypothetical impact sites of the projectile. As shown in Fig. \ref{f:wcbacc}, both the resonant re-accretion and the non-resonant collision appear in this example. Resonant mechanisms exert a strong effect within the time equivalent to $n \times P_{orb}$ where $n$ is an integer, leading to re-accretion peaks exceeding $0.1$ kg/hr. The non-resonant collisions happen as isolated small peaks between the integral times. 

\subsubsection{Kinetic energy density of the ejecta cloud} \label{s:sec32_ked} %

A first and basic description of the evolution of the kinetic energy density distribution around the system has been described in \citet{aidafateI}. The formation flying ejecta that we call the ``ejecta curtain'' can maintain for at least tens of minutes. At later times, the behavior of the sampled particles begin to diverge because of their different ejection speeds. The fraction with moderate speed will be trapped within the binary system for a relatively long period (for the $12$ cases in this section, it expands to $2-29$ days). The slow curtain that comprises this fraction of ejecta will be broken and distorted during this period, and the ejecta will be gradually cleaned via accretion/re-accretion on the binary components or ejected from the system. These behaviors are commonly found among the $12$ cases: the maximum kinetic energy density (hereafter named Max. KED) decreases rapidly as the high-speed streams that comprise the greatest flux of kinetic energy escape, and afterwards the Max. KED oscillates around a low level until the trapped ejecta are accreted/re-accreted. 

Figure \ref{f:kedcomp} illustrates the variation of Max. KED as a function of time for both materials, WCB and SFA. The curve was calculated by dividing the vicinity of Didymos into uniform meshes and taking the averaged kinetic energy over the unit volume. In reality, the kinetic energy is concentrated in the discrete ejecta pieces and therefore Max. KED only serves as a relative value rather than as the absolute estimate to measure possible injuries that a spacecraft in the vicinity would undergo if impacted. Here we adopted a grid size of $\sim 171$ m to represent the unit volume. 

\begin{center}
Figure \ref{f:kedcomp}.
\end{center}

Max. KED peaks at the excavating stage ($<1$ hr after the impact), reaching up to $2.9$ J/m$^3$ for WCB and $15.1$ J/m$^3$ for SFA, and it drops to below $10^8$ J/m$^3$ within the next $10$ hr for both materials. We assess that the results in this stage depend little on the choice of the projectile impact site, \ie\ for both materials, the data from the $6$ impact sites are in good agreement during the first $10$ hr. This is because the majority of the ejecta have large ejection speeds and the speed of the secondary contributes little to the total velocity. Afterwards, the curves of the $6$ impact sites start to diverge at a level below $10^8$ J/m$^3$, because the motion of the slow curtain depends more on the location of the impact site than the motion of high-speed ejecta. In comparison, we find that the material choice SFA leads to higher Max. KED especially during the first $10$ hr after the impact, which is consistent with the correspondingly high momentum enhancement factor (see Section \ref{s:sec31_res}), \ie\ there is more kinetic energy carried by the ejecta and a higher momentum transferred to the secondary. Moreover, the variation of Max. KED caused by the slow curtain (at times exceeding $10$ hr) behaves more periodically for WCB than for SFA for all $6$ impact sites. This can be explained, noting on one hand, that the SFA ejecta consist of smaller particles, which leads to smaller outer radii of the ejecta cloud (above the outer radius, particles will be driven off rapidly by SRP, see the discussion in \citet{aidafateI}), therefore the dust-size orbiting ejecta with high eccentricity cannot make periodic return to the neighbourhood of the binary, unlike the big fragments of WCB material; on the other hand, the SFA ejecta moving in the vicinity of the binary suffer faster orbital changes, which is also the reason why the Max. KED curve is unstable.  

\subsubsection{Dependency on the solar radiation pressure} \label{s:sec32_srp} %

\citet{jimwork} noted that particle size-sorting will occur for the ejecta accreted on the primary, and presented the distribution of ejecta projected to the celestial sphere of the Hill radius ($\sim 6$ km). The size-sorting is due to the fast rotation of the primary and the different effect of solar radiation pressure (SRP) on particles of different sizes. We further examine the SRP sensitivity of particles with different ejection speeds. Equation \ref{e:escptim} defines a baseline for the escaping particles ($EE$ \& $LE$), corresponding to the absence of gravitational perturbations from the binary components, indicating that the escape time (defined as the time when a particle moves out of the SOI) is inversely proportional to the ejection speed $v_e$. 

\begin{equation}
\label{e:escptim}
T_{esc} = \frac{R_{SOI}}{v_e},  
\end{equation}

The results derived from the simulations are consistent with Eq. \ref{e:escptim} in the high-speed range ($v_e>10$ m/s), independently on the particle size-range and material type. This suggests that the size-sorting effect may be only valid for the slow curtain that contains particles with $v_e$ below tens of cm/s, and for the high-speed curtain (if it is intercepted with the primary), the effect will be far less obvious. 

Special attention was paid to the correlations between the escape time and the ejection speed for particles of different sizes in the slow curtain. Figure \ref{f:srp1} shows the results for WCB, and Fig. \ref{f:srp2}, for SFA. The increasing effect of solar radiation pressure as the particle size decreases is remarkable, and this effect depends significantly on both the ejection speed and the impact site choice. The slow curtain, part of which is temporarily trapped in the binary system, suffers a strong separation of different ejecta sizes. Dust-size particles ($10\ \mu$m$-100\ \mu$m) are strongly accelerated by the solar radiation pressure, hence their orbits become less stable and will  be cleaned quickly. This phenomenon can be observed both in Fig. \ref{f:srp1} and Fig. \ref{f:srp2}. Overall, the small particles will escape more rapidly than the bigger ones, which correspondingly leads to a higher fraction of ``escaping'' and lower fractions of ``orbiting'' and ``escaped''. Thus, some time after the impact, the ejecta cloud evolving in the vicinity of the binary tends to be composed of larger ejecta than during the initial stage. 

\begin{center}
Figure \ref{f:srp1}.
\end{center}

\begin{center}
Figure \ref{f:srp2}.
\end{center}

Other than the correlation with the ejection speed, Fig. \ref{f:srp1} and Fig. \ref{f:srp2} also reveal some non-trivial branches of the entire distribution in a log-log scale. A tiny low-speed branch ($v_e<10$ cm/s) is found to have a short escape time ($T_{esc}<1$ day), which is widely existing for all particle sizes. We assess that it is due to the strong swing-by effect caused by close gravitational interactions with the secondary. Branches that accelerate the escape of the ejecta also appear for ejection speeds within $10$ cm/s $<v_e< 60$ cm/s, and show a significant dependency on the impact site: these gravitational-assistant escapes occur for $(90^\circ W, 0^\circ)$ and $(180^\circ, 0^\circ)$. Meanwhile, these two cases are also the ones that prove to have more ejecta exhibiting better survival advantages, \ie\ the time required for massive ejecta to move out of the SOI is longer in $(90^\circ W, 0^\circ)$ and $(180^\circ, 0^\circ)$ than in other cases. 

\section{Discussion and Conclusions} \label{s:sec4_conc}

This paper presented our study by purely numerical integrations of the evolution of the ejecta cloud from a kinetic impact on the secondary of a binary asteroid. The topic is of particular interest for the ongoing study of the space mission project AIDA, noting in particular that the artificial projectile of this mission called DART passed successfully in Phase B in June 2017. Therefore, the simulation parameters were chosen to match as best as possible the mission scenario as planned in $2022$. However, our study is also useful to indicate the regolith production by the impact process on small binary asteroid surfaces. In this study, we focused on the fate dependency of the ejecta on various parameters, including the projectile impact site, and on the various mechanisms driving the post-impact dynamics. The numerical methodology developed in our first paper \citep{aidafateI} was adopted. 

From our hierarchical grid search (Section \ref{s:sec2_atlas}), we can draw the following conclusions: 
\begin{enumerate}
\item Temporarily trapped ejecta that orbit around Didymos after the impact and ejection speeds are highly correlated. A grid search presents a closed range of ejection speeds, $4.0 < v_e < 42.0$ cm/s, beyond which the orbital behaviors of the ejecta become simplified, \ie\ residing in type $EE$/$LAP$/$LRS$.
\item Within this range ($4.0 < v_e < 42.0$ cm/s), the proportion of escaping ejecta shows a monotonous increase with increasing $v_e$, and the proportions of accreted/re-accreted ejecta and of orbiting ejecta exhibit a rule of ``ascend first and then descend'' with ejection speed. Statistically, $v_e$ around $12.0$ cm/s leads to abundant accretion/re-accretion on the binary components, and the accreting latitude on the primary proves to be correlated with the accreting speed, \eg\ re-impacts in the polar areas (latitudes above $80^\circ$ N/S) are concentrated in a narrow range around $40.0$ cm/s. 
\item The patterns of ejecta fate mapped on Didymoon's surface reveal two mechanisms that play a significant role in the evolution of the orbiting ejecta. I. The ejecta evolving on orbits in mean motion resonance with the secondary's orbit produce long-term quasi-periodic re-accretion peaks over some time, which can extend from a few days to several weeks. II. Some ejecta on non-resonant orbits produce a rapid re-accretion peak that is not recurrent, because such orbits overlap the secondary's orbit, leading to a massive re-accretion when they physically intersect. The non-resonant regime is not recurrent, but the swing-by effect occurs in both cases, which causes the motion to become chaotic and can lead ejecta with similar initial conditions to very different fates. 
\item The high spin rate of the primary leads to a relatively wide range of ejecta impact speeds. For the secondary, we find that high re-impact speeds ($>40$ cm/s) happen on the eastern hemisphere (windward) much more frequently than on the western hemisphere (leeward). 
\item The orbital distribution of the survived ejecta at the end of the two-month simulations confirm the survival advantages of the polar orbits and retrograde orbits, and further demonstrate that the quasi-polar orbits distribute on two fixed discs, leaving an ejecta-vacuum area close to the binary system. This area is found to emerge at a time posterior to the impact for any initialization of the ejecta. 
\end{enumerate}

The conclusions from the full-scale simulations (Section \ref{s:sec3_tracing}) are: 

\begin{enumerate}
\item For both material cases, an absolute majority of the ejecta exceeds the local escape speed and belongs to type $EE$. Thus a significant momentum change of the binary components seems to occur only during the excavating stage, unless the high-speed ejecta curtain sweeps across the primary directly, which is not included in our simulations. 
\item Among the $6$ hypothetical impact sites of the projectile, $(90^\circ W, 0^\circ)$ (the leeward position) and $(180^\circ, 0^\circ)$ (the inward position) favour a higher proportion of accreted/re-accreted ejecta on both components. The polar impact sites $(0^\circ, 0^\circ N/S)$ lead to longer averaged survival time of the orbiting ejecta, which agrees with the stability advantages of polar orbits. 
\item The violent period of the ejecta cloud evolution ends quickly ($<10$ hr) for all $12$ simulations, during which the high-speed streams that comprise the greatest flux of kinetic energy dissipate rapidly. Afterwards, the maximum kinetic energy density drops to and maintains a low level before the ejecta are cleaned off. 
\item The solar radiation pressure strongly accelerates the clearing of the vicinity of the binary system. In all the $12$ simulations, the ejecta are accreted/re-accreted by the components or ejected from the SOI in $1$ month, and only particles above $1$ mm in size (WCB) survive longer than $10$ days. This suggests that the particles accreted/re-accreted after this time are highly weighted towards larger ejecta. 
\end{enumerate}

One caveat of our full simulations is that they are based on an initialization of ejecta properties using scaling laws. One important weakness of those scaling laws is that they break for the low-speed ejecta \citep{crtrpara3}. Therefore, we do not claim that the low-speed ejecta production and fate from impact cratering are well accounted for in our simulations. Results from impact simulations using hydrocodes, rather than scaling laws, could be used as initial conditions of the ejecta. However, they experience the same issue, \ie\ the reliability of simulation results for the low-speed ejecta is not guaranteed when the escape speed is as low as  few cm/s, because of numerical noise and other issues whose discussion goes beyond the scope of our paper. Some efforts are under way to possibly find a solution to this issue, which will allow us to use the simulation results in our modeling of ejecta fate in a future work. Then, it would be very useful to have impact experiments performed under low-gravity conditions, focusing on the measurement of low-speed ejecta, which could then offer a possibility to improve the scaling laws and also be used to validate and therefore give more confidence to the results of numerical simulations. 

One additional issue is that low speed ejecta may not only be produced directly by the impact but may also originate from more complex mechanisms, such as regolith lift-off due to seismic shaking \citep{Garcia2015, Murdoch2017}, electrostatic levitation of fine-size particles, \etc\ Depending on the context (\ie\ high abundance of loose regolith on the whole surface), some of these mechanisms may even contribute more to the slow-speed ejecta than the cratering process itself. Future work will be performed to explore the contribution of these mechanisms. 

For this reason, in the first part of our paper, we chose to cover an arbitrarily wide range of ejection speeds, from very small values (below Didymoon's escape speed) to high ones, independently of their plausibility. This strategy is not meant to represent a real case, but rather to offer a global view of the behavior of ejecta as a function of their ejection speed. In other words, if any of the ejecta is ejected with the assumed speed, it will have the behavior calculated with our model, which is already a useful information, and the only one we can provide, given the issues indicated above. 

We plan to continue our investigations, accounting for the above-mentioned issues, in various asteroid environments in order to better understand the contribution of ejecta to asteroid regolith and to support space mission studies. For the former, we will explore various kinds of asteroids, accounting for their different sizes (gravity), shapes, spin rates, possible mechanical properties, binarity, \etc\ For the latter, we will keep studying this process in the context of the AIDA project as it evolves and the Japanese Hayabusa2 mission, which carries a Small Carry-on Impactor \citep{Arakawa2017} that will perform the impact of a 2 kg copper projectile at 2 km/s on the asteroid Ryugu surface and which needs information on ejecta fate to determine the positioning of the spacecraft during the experiment and on the distribution of re-accreted ejecta for the sampling planned after this experiment.

\newpage
\section*{Acknowledgements} \label{s:ackndgmt}
\addcontentsline{toc}{section}{\nameref{s:ackndgmt}}

Y.Y. and P.M. acknowledge the support from the NEOShield-2 project, funded under the European Commission'™s Horizon 2020 research and innovation programme under grant agreement No. 640351. P.M. acknowledge support from ESA and CNES. Y.Y. and P.M. thank Stephen R. Schwartz for the discussion on the conception of this work, and Shantanu P. Naidu as well as Lance A.M. Benner for providing the shape model of the primary of Didymos. The reference parameters of Didymos system were provided by the ESA-AIM Team. The simulations were performed partly using the cluster Licallo at l'Observatoire de la C\^ote d'Azur, CNRS (France) and partly using the cluster Tianhe-$2$ located in the Chinese National Supercomputer Center (China). 

\newpage
\section*{References} \label{s:sec6_refs}
\addcontentsline{toc}{section}{\nameref{s:sec6_refs}}
\def\refname{}

\newpage

\section*{Appendix A: Binary Initial Configurations and Surface Mapping} \label{s:appdxA}
\addcontentsline{toc}{section}{\nameref{s:appdxA}}

The initial configuration of the binary components is defined in the reference model of $65803$ Didymos of the AIDA-AIM team, which is also exposed in \citet{aimsci}.  The spin rate of the primary, the period of the mutual orbit $P_{orb}$, the diameter ratio between the components and the length of their separation are measured directly by remote observations. Thus in our model, these measured properties are fixed and the other physical and dynamical parameters are assumed to make the whole system's properties consistent \citep{aimsci, aidafateI}. In order to restore the heliocentric orientation of Didymos, we first adopted the retrograde solution of the mutual orbit by \citet{schpra}, \ie\ $\left ( 310^\circ, -84^\circ \right )$ in the ecliptic coordinate system. Then assuming the orbital plane is perpendicular to the rotational axis of the primary (assumed to be the principal axis that maximizes the moment of inertia), we determined the orientation of the binary system with respect to the heliocentric frame (see Fig \ref{f:binorn}). The positions and attitudes of the binary components were represented in the orbital translational frame, which originates at the center of mass of the binary and with axes consistent with the heliocentric ecliptic J$2000$ (the $0^\circ$ ecliptic longitude points to J$2000$ mean equinox, and the system has a right-hand convention). 

\begin{center}
Figure \ref{f:binorn}.
\end{center}

The relative rotational states of the primary and the secondary will not be known with high accuracy prior to the rendezvous of a spacecraft, however as found for other binary systems and natural satellites, it is likely that its rotational state is tidally locked because of internal dissipations. This technically enables us to set up the system initial configurations by aligning the body frames of the primary and the secondary. Defining the $x$-, $y$-, $z$-axes of the body-fixed frame as the minimum, medium, maximum principal axes of inertia, respectively, we assume a locked phase for the initial relative rotation, \ie\ the long axis $x$ of Didymoon is oriented towards the primary's center of mass, and the rotational axis $y$ is parallel with that of the primary, but with a spinning period equal to $P_{orb}$. Furthermore, we assume that the primary's $x$-axis is aligned to that of the secondary at the impact time (which is practically unpredictable but has little influence thanks to the axial symmetry of the primary). Then the initial configuration of the binary components is set up as shown in Fig. \ref{f:binorn}. 

The instantaneous solar position was derived from the Solar System ephemeris extrapolation to the impact time. The arrival time of the AIDA impactor DART is estimated to be late September / early October $2022$, and we chose as the baseline impact time the perigee time of Didymos as $2022$/$10$/$04$ at $09$:$48$:$00$ UTC. The solar position is calculated to be $\sim 193^\circ$ in ecliptic longitude. A full eclipse phase of Didymoon is shown in Fig. \ref{f:binorn}. For low-speed and dust-size ejecta, given the high influence of solar radiation pressure on their evolution, the assumed position of the binary system with respect to the Sun plays a big role \citep{aidafateI}, and it should be considered as a key factor when examining the dynamical outcomes of this fraction of ejecta. 

The geographic coordinate systems of the binary component surfaces are defined based on the body frames. For Didymoon, we define $xy$ as the equatorial plane, and $+x$ points to $0^\circ$ longitude.  The longitude (denoted by $\lambda$) ranges from $-180^\circ$ westward to $+180^\circ$ eastward, and $+y$ defines the eastern hemisphere (facing the flying direction of Didymoon). Then using the right-hand convention, $+z$ points to the North Pole, and the latitude (denoted by $\phi$) ranges from $-90^\circ$ southward to $+90^\circ$ northward (measured from the equator). The longitude and latitude on the surface of the primary are defined in the same way as for the Didymoon. 

\newpage

\section*{Appendix B: The Sampling Scheme} \label{s:appdxB}
\addcontentsline{toc}{section}{\nameref{s:appdxB}}

\setcounter{equation}{0}    					
\renewcommand{\theequation}{B-\arabic{equation}} 	

For the discretisation of ejecta defined by scaling laws, a piecewise scheme can unrealistically break the smoothness of the ejecta medium in the subsequent evolution. Here we propose an improved scheme from that of \citet{aidafateI} purely based on a uniform random number generator, and relying on no divisions of the sampled ranges. Given a full ejecta size range $[d_0, d_1]$ ($d_l \le d_0 < d_1 \le d_u$) and a full interval of radial distance from a crater's center $[x_0, x_1]$ ($n_1 a \le x_0 < x_1 \le n_2 R$), the reference number $N_r$ (Eq. (\ref{e:pwlw})) is limited by the total ejecta mass that resides in $[d_0, d_1] \times [x_0, x_1]$: 

\begin{equation}
\label{e:nrlmt}
-\int_{d_0}^{d_1} \rho_s \frac{4 \pi}{3} \left( \frac{d}{2} \right)^3 N_r t d^{t-1} \textup{d} d \le M \left ( < x_1 \right ) - M \left ( < x_0 \right ), 
\end{equation}

where $\rho_s$ indicates the bulk density of individual grains, and $t$ indicates the slope of the power law of their cumulative size distribution (its value is $2.8$ in Eq.~(\ref{e:pwlw})). The equality of Eq. (\ref{e:nrlmt}) holds if $d_0 = d_l$ and $d_1 = d_u$, \ie\

\begin{equation}
\label{e:nrmax}
N_{r, Max} = - \frac{9 k m \rho (t+3) (x_1^3-x_0^3)}{2 \pi^2 \delta a^3 t \rho_s (d_u^{t+3}-d_l^{t+3})}. 
\end{equation}

As $N_r$ is specified, the sample size $N_s$ over the parameter space $[d_0, d_1] \times [x_0, x_1]$ is determined by

\begin{equation}
\label{e:nsval}
N_s = N_{r, Max} (d_0^t-d_1^t). 
\end{equation}

We start with a uniform distribution over $[d_0, d_1] \times [x_0, x_1]$ based on a random number generator, \eg\ Lehmer algorithm. $N_s$ random numbers are created within $[d_0, d_1]$ defined as set $\mathscr{D}$ of numbers. Then, an inversible transformation (Eq.~(\ref{e:gform}) is applied to $\mathscr{D}$. 

\begin{equation}
\label{e:gform}
d' = g(d) = \left[ \frac{d-d_0}{d_1-d_0} (d_1^t-d_0^t) + d_0^t \right]^{1/t}. 
\end{equation}

It is easy to show that $g$ is monotonic and satisfies $g(d_0) = d_0$ and $g(d_1) = d_1$. Therefore the transformed set $g(\mathscr{D})$ remains within $[d_0, d_1]$ and follows the power law distribution defined by Eq.~(\ref{e:pwlw}).

Likewise, for the range of radial distance $[x_0, x_1]$, we first generate $N_s$ random numbers, denoted as set $\mathscr{X}$, and assuming the size distribution does not depend on the radial distance, the final distribution satisfies:

\begin{equation}
\label{e:MNrel}
\frac{\textup{d}N}{\textup{d}x} \propto \frac{\textup{d}M}{\textup{d}x} = m \frac{9 k}{4 \pi} \frac{\rho}{\delta} \frac{x^2}{a^3}. 
\end{equation}

We introduce an inversible transformation (Eq.~(\ref{e:fform})) and apply it to $\mathscr{X}$,

\begin{equation}
\label{e:fform}
x' = f(x) = \left[ \frac{x-x_0}{x_1-x_0} (x_1^3-x_0^3) + x_0^3 \right]^{1/3}. 
\end{equation}

Then we can verify that $f$ is monotonic and satisfies $f(x_0) = x_0$, $f(x_1) = x_1$. The transformed set $f(\mathscr{X})$ remains within $[x_0, x_1]$ and follows the scaling law distribution defined by Eq.~(\ref{e:sclM}). $g(\mathscr{D})$ and $f(\mathscr{X})$ make up a random sample of size $N_s$ over $[d_0, d_1] \times [x_0, x_1]$, which inherently follows the power law and scaling law distributions. 

\newpage 
\section*{Figure Captions} \label{s:figcaps}
\addcontentsline{toc}{section}{\nameref{s:figcaps}}

\begin{figure}[h]
\centering
\includegraphics[width=1.0\textwidth] {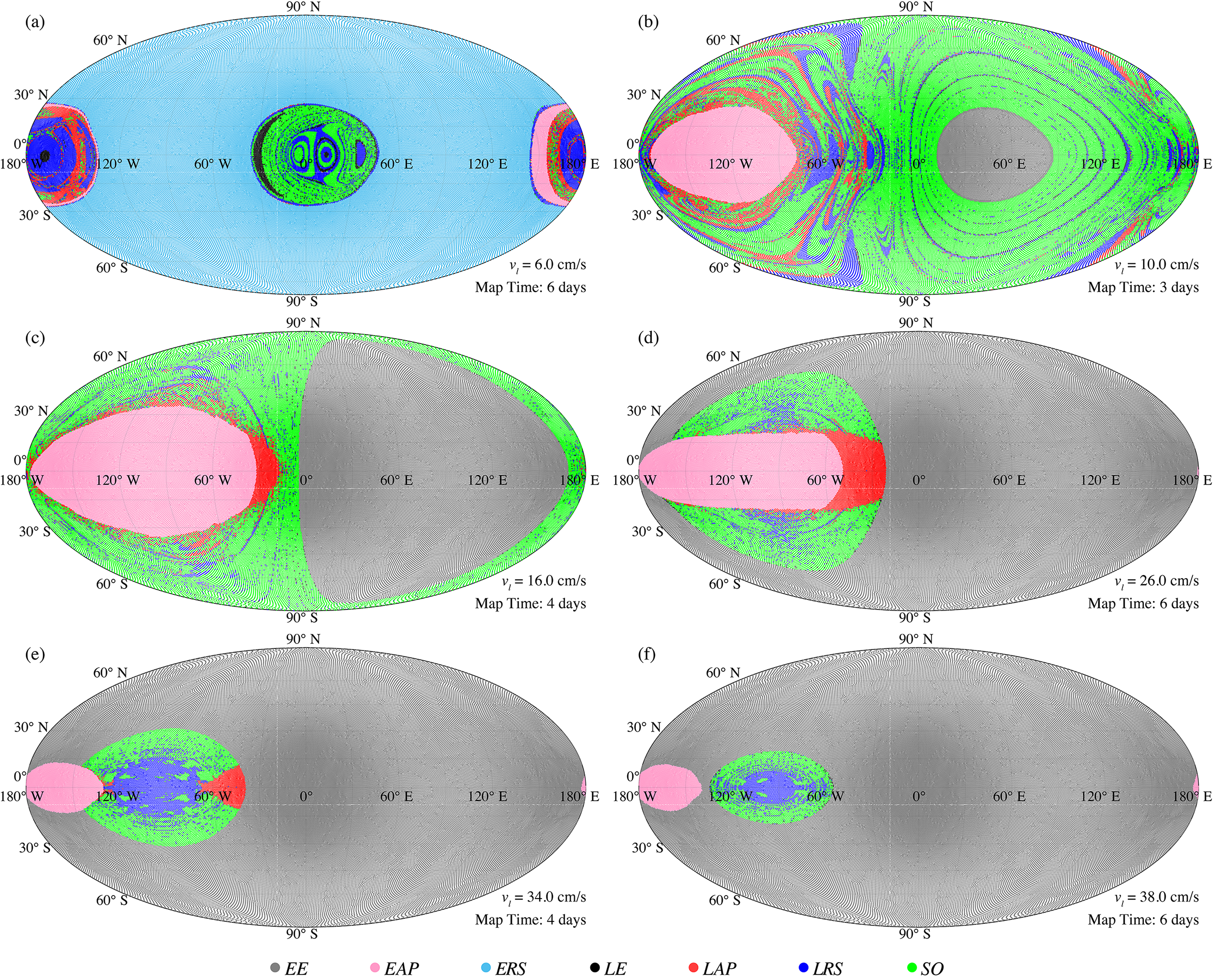} 
\caption{}
\label{f:fateatlas}
\end{figure}

\begin{figure}[h]
\centering
\includegraphics[width=0.55\textwidth] {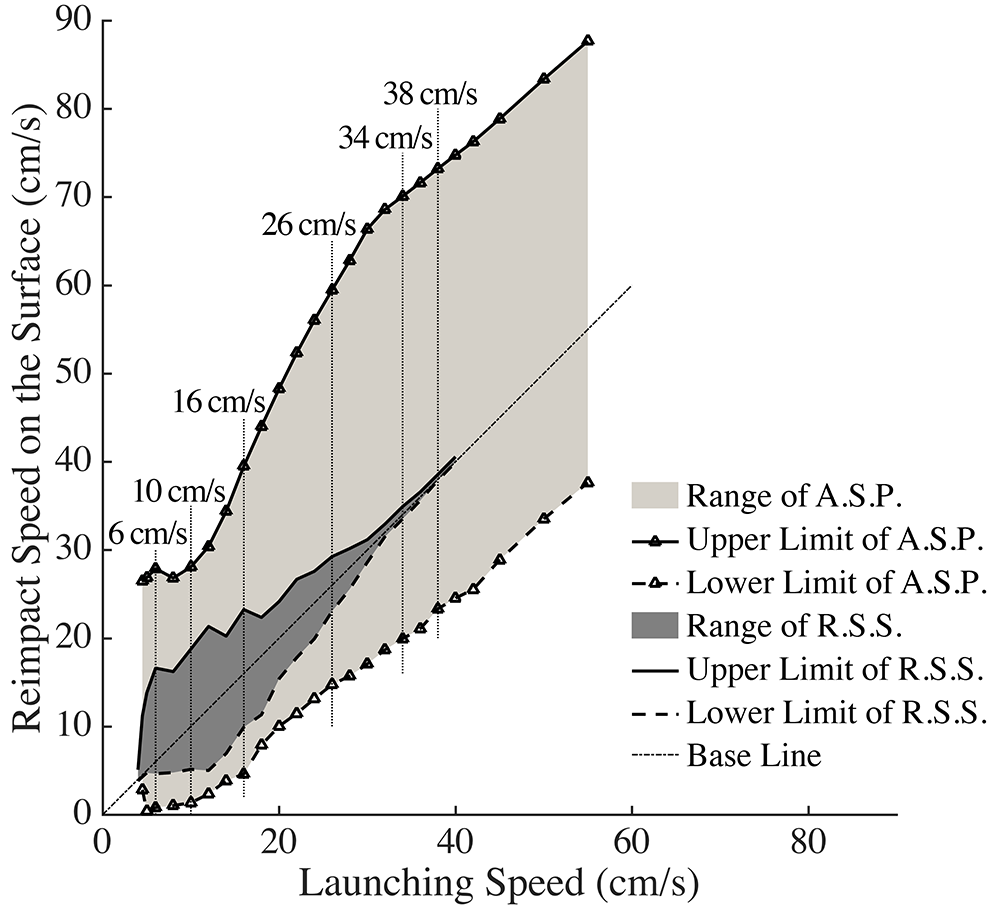} 
\caption{}
\label{f:reimpact}
\end{figure}

\begin{figure}[h]
\centering
\subfigure{
\label{}
\includegraphics[width=0.47\textwidth] {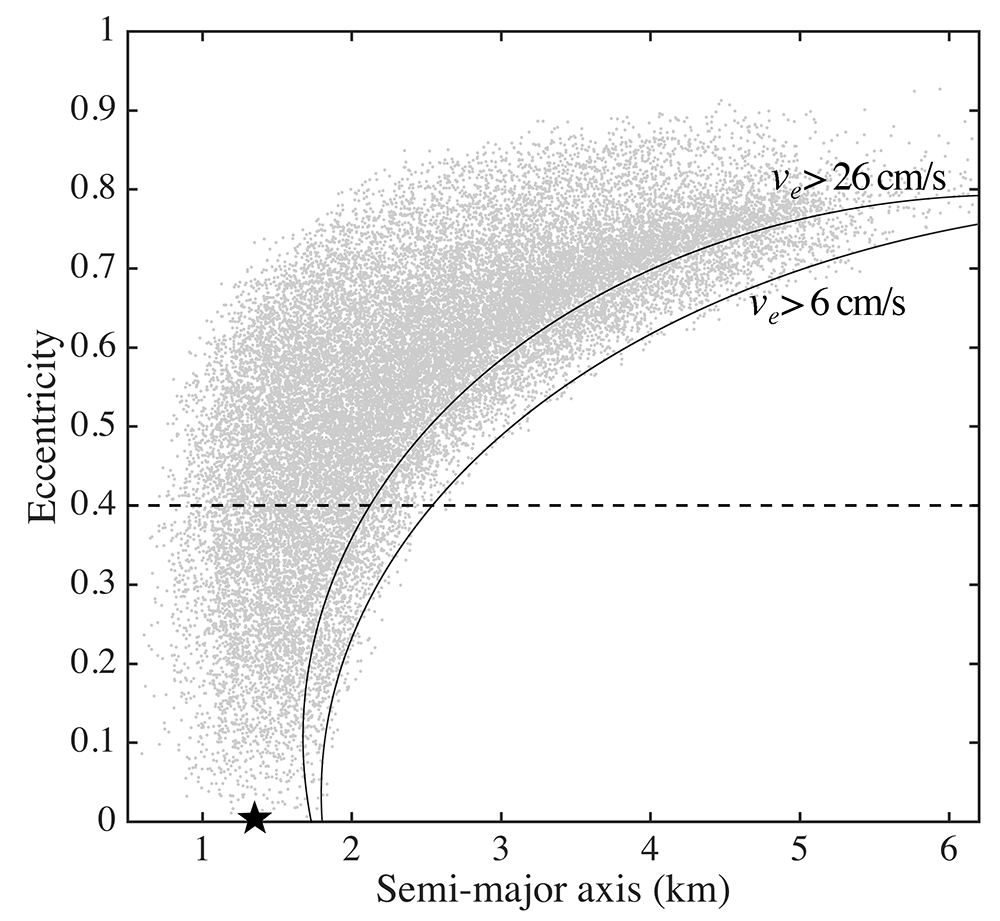} }
\subfigure{
\label{}
\includegraphics[width=0.47\textwidth] {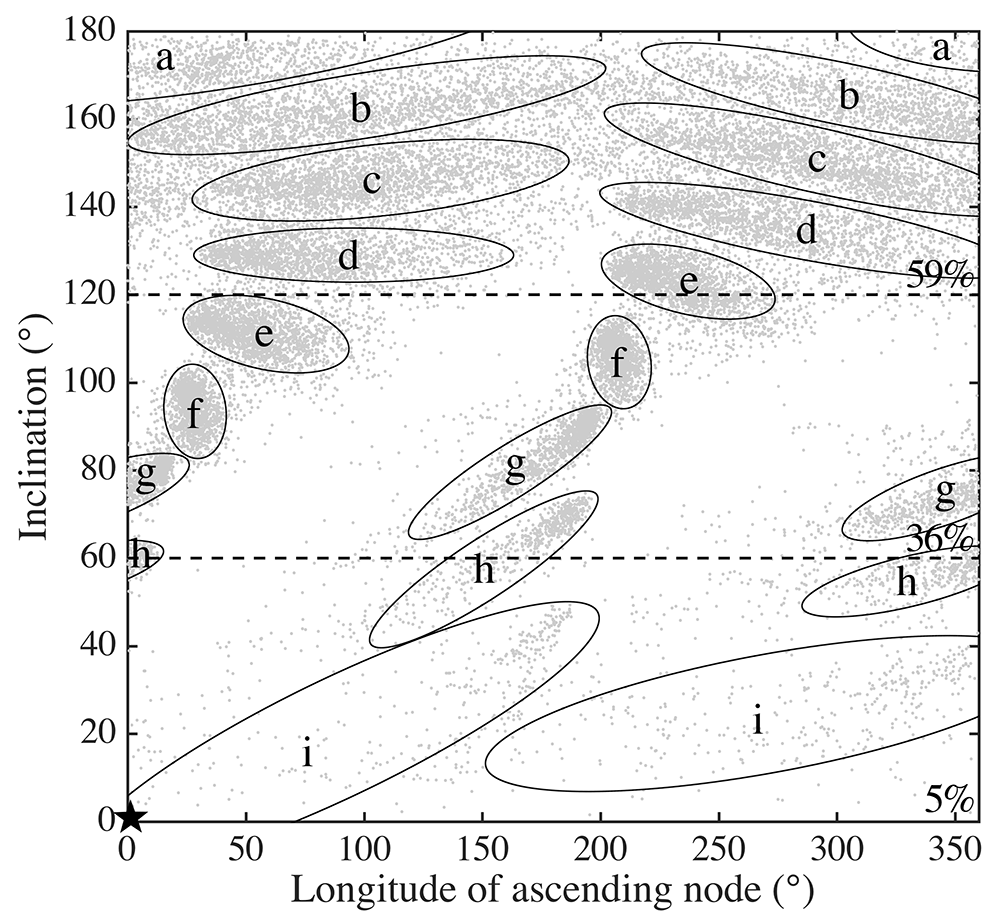} }
\caption{}
\label{f:orbdist}
\end{figure}

\begin{figure}[h]
\centering
\includegraphics[width=0.55\textwidth] {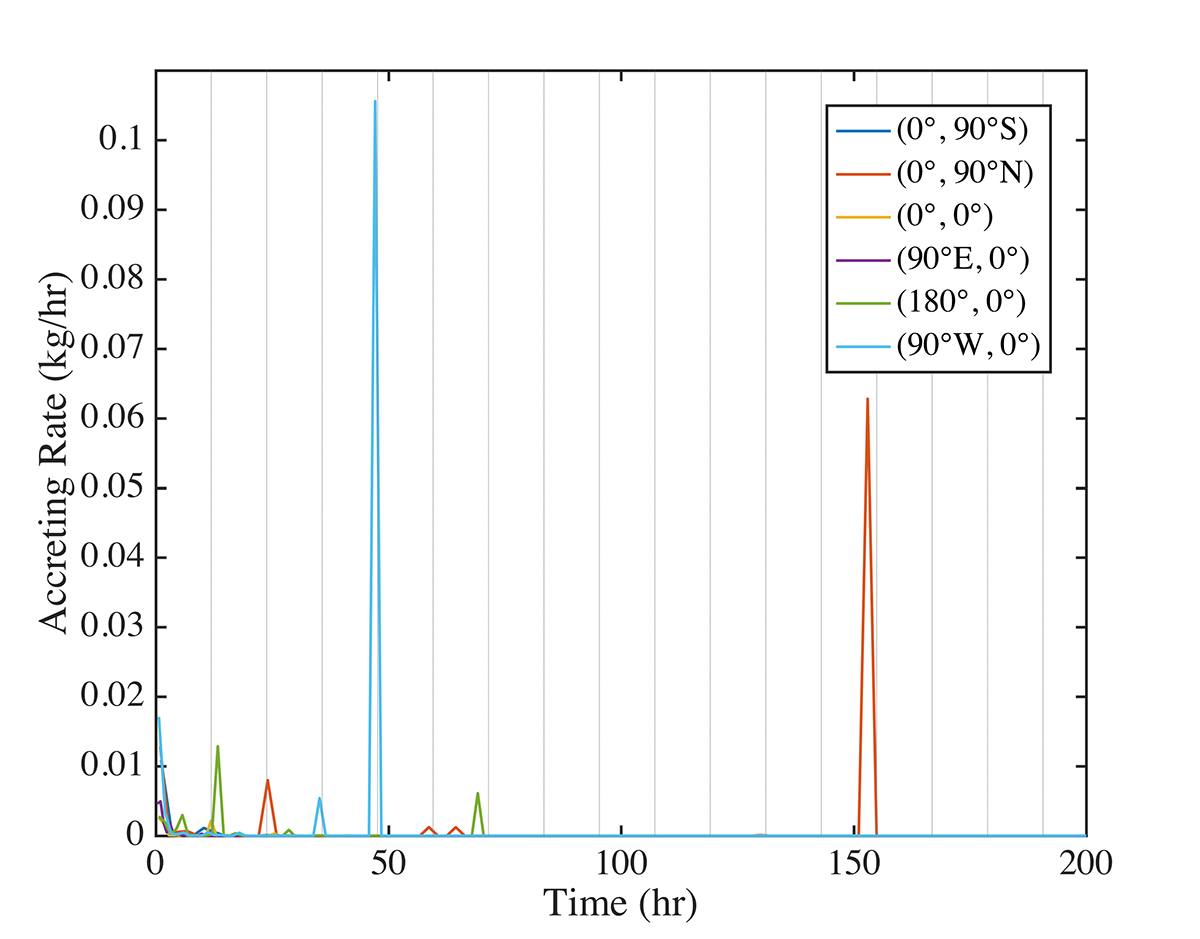}
\caption{}
\label{f:wcbacc}
\end{figure}

\begin{figure}[h]
\centering
\includegraphics[width=0.52\textwidth] {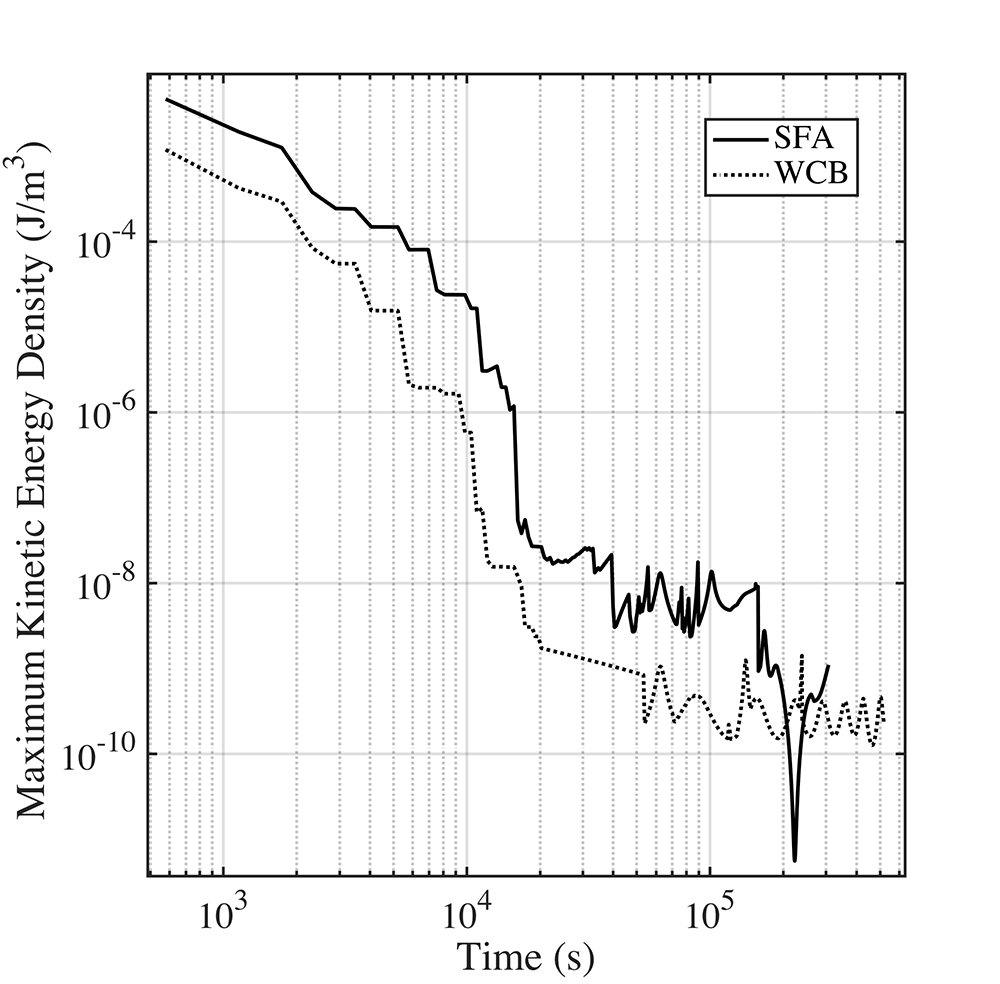}
\caption{}
\label{f:kedcomp}
\end{figure}

\begin{figure}[h]
\centering
\includegraphics[width=0.98\textwidth] {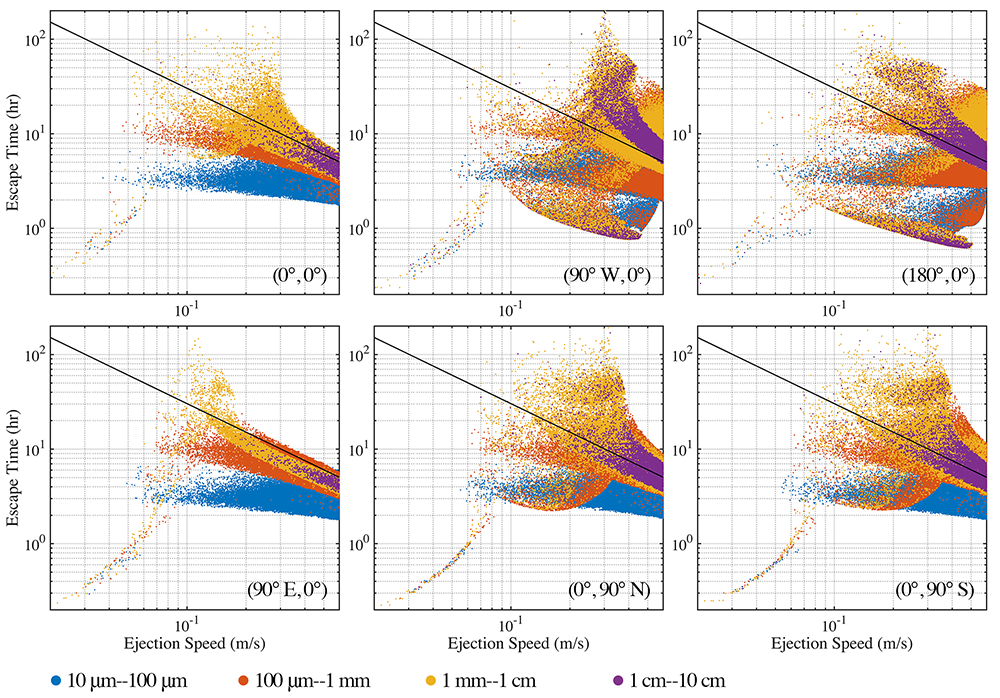}
\caption{}
\label{f:srp1}
\end{figure}

\begin{figure}[h]
\centering
\includegraphics[width=0.98\textwidth] {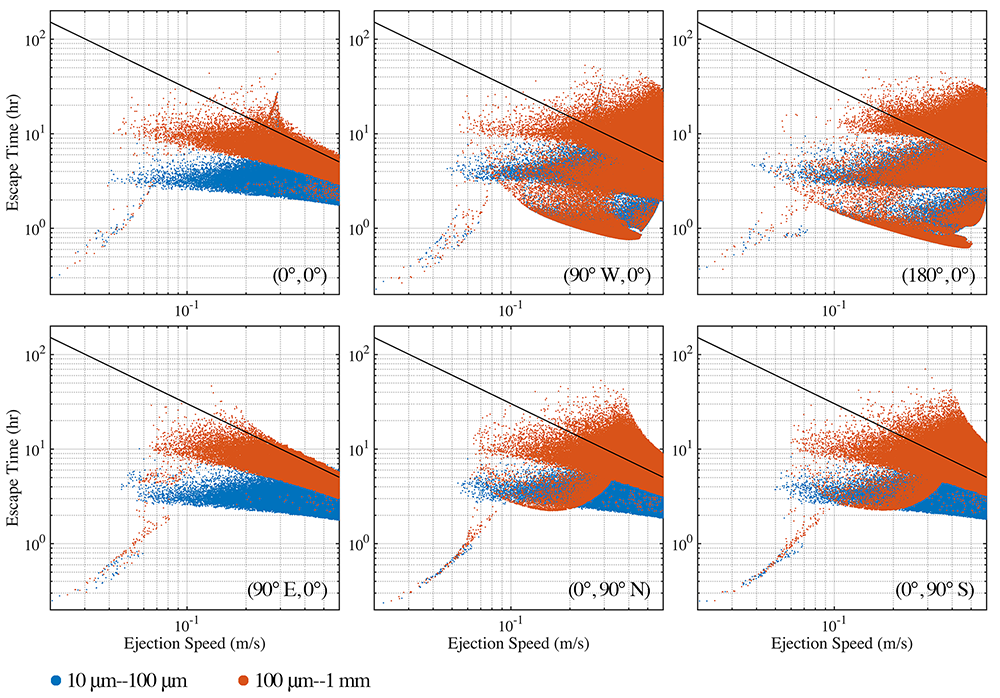}
\caption{}
\label{f:srp2}
\end{figure}

\begin{description}
\item[Figure \ref{f:fateatlas}:] { The dynamical fates of the ejecta particles of different ejection speeds from Didymoon's surface. The maps show the distribution of the $7$ types of dynamical fates pictured against the launching sites of the sampled particles, each for a unified ejection speed as (a) $6.0$ cm/s (b) $10.0$ cm/s (c) $16.0$ cm/s (d) $26.0$ cm/s (e) $34.0$ cm/s and (f) $38.0$ cm/s. }
\item[Figure \ref{f:reimpact}:] { The ranges of accreted speeds on the primary (A.S.P.) and re-accreted speeds on the secondary (R.S.S) in terms of the simulation results (as a function of the ejection speed). The accreted and re-accreted speeds are defined in the local frames with respect to the surfaces of the binary components, respectively. The boundary curves indicate the upper and lower limits of the corresponding range. The $6$ $v_e$ values from Fig. \ref{f:fateatlas} are marked with dotted lines for comparison. }
\item[Figure \ref{f:orbdist}:] { The instantaneous Keplerian orbits of the survived ejected particles at $T_f$, represented in the semi-major axis vs. eccentricity plane (left) and the longitude of ascending node vs. inclination plane (right). The solid star shows the position of the orbit of the secondary. The solid lines of the left plot indicate the upper boundaries of the semi-major axis for particles with different ejection speeds. The labeled ellipses in the right plot mark the clumps of survived orbits for different ejection speeds: a=$6$ cm/s, b=$10$ cm/s, c=$14$ cm/s, d=$18$ cm/s, e=$22$ cm/s, f=$26$ cm/s, g=$30$ cm/s, h=$34$ cm/s and i=$38$ cm/s. The dashed lines in the right plot mark out three ranges of the inclination, and their corresponding percentages of surviving ejecta are labeled in the bottom right. }
\item[Figure \ref{f:wcbacc}:] { The re-accreting rate on the secondary's surface for material WCB. The curves in colours indicate results for the ejecta cloud from the $6$ chosen impact sites of the projectile indicated on the plot. The time between two vertical bars corresponds to $P_{orb}$. }
\item[Figure \ref{f:kedcomp}:] { Maximum kinetic energy density carried by the ejecta as a function of time. Two curves are shown for the material case WCB $(0^\circ, 90^\circ N)$ and SFA $(0^\circ, 90^\circ N)$, respectively. }
\item[Figure \ref{f:srp1}:] { Evolution of the low-speed curtain of the ejecta cloud ($<60$ cm/s) from the $6$ impact sites of the projectile using material WCB, projected in the escape time vs. ejection speed plane. Particles of different size ranges are marked by different colours as the legend indicates, and the solid reference line is defined by Eq. \ref{e:escptim}. }
\item[Figure \ref{f:srp2}:] {Same as Fig. \ref{f:srp1} for material SFA. }
\item[Figure \ref{f:binorn}:] { Initial configuration of Didymos system oriented in the celestial ecliptic reference frame. The nominal orbital pole and the initial spin poles of both components are all assumed to be aligned at the AIDA/DART impact time, along $\left ( 310^\circ, -84^\circ \right )$ in the ecliptic reference frame. }
\end{description}


\newpage 

\newpage 

\newpage 

\newpage 

\newpage 

\newpage 

\newpage 

\setcounter{figure}{0}    					
\renewcommand{\thefigure}{A\arabic{figure}} 	

\newpage 
\begin{figure}[h]
\centering
\includegraphics[width=0.75\textwidth] {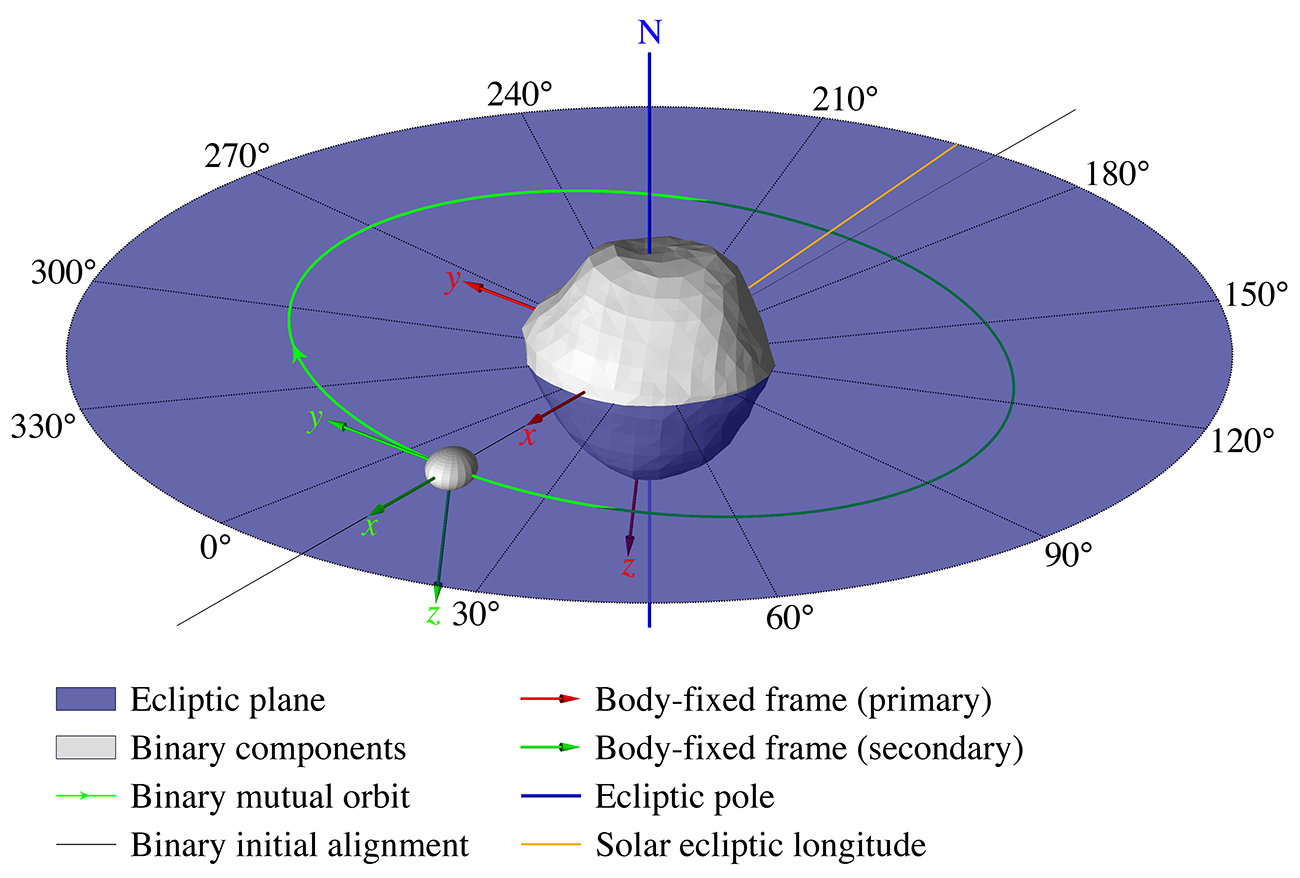} 
\caption{}
\label{f:binorn}
\end{figure}

\clearpage

\begin{table}[ht]
\centering
\caption[]{Ejecta sampling strategy and tracer particle parameters. }
\label{t:sampara1}
\vspace{0.2in}
\begin{threeparttable}
\begin{tabular}{l l}
\hline
Sampling parameters: 	& \\
\hline
\ \ \ Sample size 		& $95,486$ particles \\
\ \ \ Sample range 		& Global surface of Didymoon \\
\ \ \ Sample distribution 	& Uniform within the sampling range$^*$ \\
\ \ \ Launching direction 	& Local normal (outwards) \\
\hline
Tracer particle properties: 	& \\
\hline
\ \ \ Diameter 			& $10.0$ cm \\
\ \ \ Solid density 		& $3.4$ g/cc \\
\ \ \ Reflection rate		& $0.15$ \\ 
\hline
\end{tabular}
\begin{tablenotes}
\item[*] The distribution is uniform in terms of particle numbers per unit area within the sampling range. The initial coordinates of the tracer particles are derived based on a previous surface division using triangular meshes. 
\end{tablenotes}
\end{threeparttable}
\end{table}

\clearpage

\begin{table}
\caption[]{Statistics of the ejecta fates according to the ``$68$-$95$-$99.7$'' rule, together with the ultimate states at the termination time. Proportions are derived from simulations within a duration of $1440$ hr after the ejecta release from Didymoon. }
\label{t:glbres}
\vspace{0.2in}
\resizebox{\textwidth}{!}{
\begin{threeparttable}
\begin{tabular}{c|cccc|cccc|cccc|ccccc}
\hline
\multirow{2}{*}{$v_e$ (cm/s)} & \multirow{2}{*}{$T_{1\sigma}$ (hr)} & \multicolumn{3}{c|}{$1\sigma$.Comp. ($68\%$)} & \multirow{2}{*}{$T_{2\sigma}$ (hr)} & \multicolumn{3}{c|}{$2\sigma$.Comp. ($95\%$)} & \multirow{2}{*}{$T_{3\sigma}$ (hr)} & \multicolumn{3}{c|}{$3\sigma$.Comp. ($99.7\%$)} & \multirow{2}{*}{$T_f$ (hr)} & \multicolumn{4}{c}{Term.Comp.} \\
\cline{3-5} \cline{7-9} \cline{11-13} \cline{15-18}
 & & Acc.P. & Acc.S. & Escp. & & Acc.P. & Acc.S. & Escp. & & Acc.P. & Acc.S. & Escp. & & Acc.P. & Acc.S. & Escp. & Orb. \\ 
\hline
$4.0$   & $0.71$     & $0\%$        & $68\%$      & $0\%$        & $1.02$       	& $0\%$ 		& $95\%$      & $0\%$        & $1.27$       & $0\%$          & $99.7\%$   & $0\%$        & $1.35$ 	& $0\%$        & $100\%$     & $0\%$       & $0\%$ 		\\ 
$4.5$   & $0.87$     & $0\%$        & $68\%$      & $0\%$        & $1.47$       	& $0\%$ 		& $95\%$      & $0\%$        & $2.76$ 	      & $0\%$          & $99.7\%$   & $0\%$        & $99.36$ 	& $0.06\%$   & $99.94\%$  & $0\%$       & $0\%$ 		\\ 
$5.0$   & $1.97$     & $0\%$        & $68\%$      & $0\%$        & $4.45$       	& $0.17\%$ 	& $94.83\%$ & $0\%$        & $1282.22$ & $3.09\%$     & $95.58\%$ & $1.03\%$   & $1440$ 	& $3.09\%$   & $95.59\%$  & $1.07\%$  & $0.25\%$ 	\\ 
$6.0$   & $1.95$     & $0\%$        & $68\%$      & $0\%$        & $360.56$   	& $6.05\%$ 	& $85.94\%$ & $3.01\%$   & -- 		      & -- 	             & -- 		 & -- 		      & $1440$ 	& $6.26\%$   & $87.80\%$  & $5.16\%$  & $0.78\%$ 	\\ 
$8.0$   & $132.22$ & $13.25\%$ & $51.38\%$ & $3.37\%$   & $906.39$   	& $16.85\%$ 	& $64.85\%$ & $13.30\%$ & -- 		      & -- 	             & -- 		 & -- 		      & $1440$ 	& $17.11\%$ & $65.98\%$  & $14.93\%$ & $1.98\%$ 	\\ 
$10.0$ & $264.44$ & $24.24\%$ & $28.45\%$ & $15.31\%$ & $1216.39$ 	& $28.60\%$	& $39.76\%$ & $26.64\%$ & -- 		      & -- 	             & -- 		 & -- 		      & $1440$ 	& $28.76\%$ & $40.28\%$  & $27.38\%$ & $3.58\%$ 	\\ 
$12.0$ & $238.00$ & $26.51\%$ & $13.91\%$ & $27.58\%$ & --               	& -- 			& -- 	             & -- 		 & -- 		      & -- 	             & -- 		 & -- 		      & $1440$ 	& $31.56\%$ & $24.50\%$  & $38.78\%$ & $5.16\%$ 	\\ 
$14.0$ & $144.25$ & $25.03\%$ & $6.44\%$   & $36.53\%$ & --               	& -- 			& -- 	             & -- 		 & -- 		      & -- 	             & -- 		 & -- 		      & $1440$ 	& $30.94\%$ & $16.80\%$  & $46.07\%$ & $6.19\%$ 	\\ 
$16.0$ & $58.42$   & $22.09\%$ & $1.78\%$   & $44.13\%$ & --               	& -- 			& -- 	             & -- 		 & -- 		      & -- 	             & -- 		 & -- 		      & $1440$ 	& $29.36\%$ & $12.99\%$  & $51.73\%$ & $5.92\%$ 	\\ 
$18.0$ & $25.44$   & $20.70\%$ & $0.58\%$   & $46.72\%$ & --               	& -- 			& --	             & -- 		 & --     	      & -- 	             & -- 		 & -- 		      & $1440$ 	& $27.13\%$ & $10.48\%$  & $57.09\%$ & $5.30\%$ 	\\ 
$20.0$ & $19.79$   & $19.66\%$ & $0.29\%$   & $48.05\%$ & $1370.28$ & $24.73\%$ 	& $8.60\%$   & $61.67\%$ & --     	      & -- 	             & -- 		 & -- 		      & $1440$ 	& $24.78\%$ & $8.70\%$    & $61.78\%$ & $4.74\%$ 	\\ 
$22.0$ & $16.83$   & $18.17\%$ & $0.23\%$   & $49.60\%$ & $1293.33$ 	& $22.32\%$ 	& $7.30\%$   & $65.38\%$ & -- 		      & -- 	             & -- 		 & -- 		      & $1440$ 	& $22.40\%$ & $7.50\%$    & $65.55\%$ & $4.55\%$ 	\\ 
$24.0$ & $15.19$   & $16.21\%$ & $0.20\%$   & $51.59\%$ & $1098.61$ 	& $19.69\%$ 	& $6.25\%$   & $69.06\%$ & -- 		      & -- 	             & -- 		 & -- 		      & $1440$ 	& $19.84\%$ & $6.75\%$    & $69.34\%$ & $4.07\%$ 	\\ 
$26.0$ & $13.94$   & $14.04\%$ & $0.19\%$   & $53.77\%$ & $935.83$   	& $17.14\%$ 	& $5.52\%$   & $72.34\%$ & -- 		      & -- 	             & -- 		 & -- 		      & $1440$ 	& $17.32\%$ & $6.25\%$    & $72.66\%$ & $3.77\%$ 	\\ 
$28.0$ & $12.82$   & $11.60\%$ & $0.19\%$   & $56.21\%$ & $745.28$   	& $14.39\%$ 	& $5.06\%$   & $75.55\%$ & -- 		      & -- 	             & -- 		 & -- 		      & $1440$ 	& $14.63\%$ & $6.07\%$    & $75.97\%$ & $3.33\%$ 	\\ 
$30.0$ & $12.34$   & $8.71\%$   & $0.22\%$   & $59.07\%$ & $507.22$   	& $11.38\%$ 	& $5.00\%$   & $78.62\%$ & -- 		      & -- 	             & -- 		 & -- 		      & $1440$ 	& $11.56\%$ & $6.50\%$    & $79.08\%$ & $2.86\%$ 	\\ 
$32.0$ & $11.86$   & $5.07\%$   & $0.29\%$   & $62.64\%$ & $329.44$   	& $7.11\%$ 	& $6.15\%$   & $81.74\%$ & -- 		      & -- 	             & -- 		 & -- 		      & $1440$ 	& $7.33\%$   & $8.22\%$    & $82.21\%$ & $2.24\%$ 	\\ 
$34.0$ & $11.18$   & $3.61\%$   & $0.26\%$   & $64.13\%$ & $189.92$   	& $4.90\%$ 	& $5.32\%$   & $84.78\%$ & -- 		      & -- 	             & -- 		 & -- 		      & $1440$ 	& $5.16\%$   & $8.12\%$    & $85.33\%$ & $1.39\%$ 	\\ 
$36.0$ & $10.26$   & $3.16\%$   & $0\%$        & $64.84\%$ & $141.83$   	& $3.40\%$ 	& $3.51\%$   & $88.09\%$ & -- 		      & -- 	             & -- 		 & -- 		      & $1440$ 	& $3.64\%$   & $6.59\%$    & $88.62\%$ & $1.15\%$ 	\\ 
$38.0$ & $9.62$     & $2.85\%$   & $0\%$        & $65.15\%$ & $40.56$   	& $2.85\%$ 	& $0.63\%$   & $91.52\%$ & -- 		      & -- 	             & -- 		 & -- 		      & $1440$ 	& $2.85\%$   & $4.23\%$    & $92.25\%$ & $0.67\%$ 	\\ 
$40.0$ & $8.89$     & $2.62\%$   & $0\%$        & $65.38\%$ & $22.00$     	& $2.62\%$ 	& $0\%$        & $92.38\%$ & $782.22$   & $2.62\%$     & $1.41\%$   & $95.67\%$  & $1440$ 	& $2.62\%$   & $1.51\%$    & $95.72\%$ & $0.15\%$ 	\\ 
$42.0$ & $8.41$     & $2.43\%$   & $0\%$        & $65.56\%$ & $16.83$     	& $2.43\%$ 	& $0\%$        & $92.57\%$ & $25.24$     & $2.43\%$     & $0\%$        & $97.27\%$  & $27.16$ 	& $2.43\%$   & $0\%$  	  & $97.57\%$ & $0\%$ 		\\ 
$45.0$ & $7.69$     & $2.25\%$   & $0\%$        & $65.75\%$ & $13.22$     	& $2.25\%$ 	& $0\%$        & $92.75\%$ & $16.35$     & $2.25\%$     & $0\%$        & $97.45\%$  & $17.23$ 	& $2.25\%$   & $0\%$  	  & $97.75\%$ & $0\%$ 		\\ 
$50.0$ & $6.73$     & $2.03\%$   & $0\%$        & $65.97\%$ & $10.22$     	& $2.03\%$ 	& $0\%$        & $92.97\%$ & $12.02$     & $2.03\%$     & $0\%$        & $97.67\%$  & $12.42$ 	& $2.03\%$   & $0\%$  	  & $97.97\%$ & $0\%$ 		\\ 
$55.0$ & $6.02$     & $1.88\%$   & $0\%$        & $66.12\%$ & $8.44$     	& $1.88\%$ 	& $0\%$        & $93.12\%$ & $9.47$       & $1.88\%$     & $0\%$        & $97.82\%$  & $10.08$ 	& $1.88\%$   & $0\%$  	  & $98.12\%$ & $0\%$ 		\\ 
\hline
\end{tabular}
\begin{tablenotes}
\item[*] -- indicates that the $2\sigma$/$3\sigma$ control limits are not applicable because the required residual percentages of orbiting particles are not satisfied at the termination time. 
\end{tablenotes}
\end{threeparttable}
}
\end{table}

\clearpage

\begin{table}[!ht]
\centering
\caption[]{Impact ejecta scaling parameters of weakly cemented basalt (WCB) and sand/fly ash (SFA).}
\label{t:materials}
\vspace{0.2in}
\begin{tabular}{c c c c c c c c c c c c }
\hline
		& $Y$ (MPa) & $\rho$ (g/cc) & $\mu$ & $H_2$  & $C_1$  & $k$    & $p$    & $\nu$ & $n_1$ & $n_2$ \\
\hline
 WCB 	& $0.45$       & $2.6$            & $0.46$ & $0.38$ & $0.18$ & $0.3$ & $0.3$ & $0.4$ & $1.2$ & $1.0$ \\
 SFA 	& $0.004$     & $1.5$            & $0.4$   & $0.4$   & $0.55$ & $0.3$ & $0.3$ & $0.4$ & $1.2$ & $1.0$ \\
\hline
\end{tabular}
\end{table}

\clearpage

\begin{table}[!ht]
\centering
\caption[]{Summary of the discretised cratering outcomes derived from the scaling and distribution laws using the material parameters of weakly cemented basalt and sand/fly ash. The extremes of the grain size range as well as the crater radius and the size composition of the ejecta are calculated for both materials. }
\label{t:discret}
\vspace{0.2in}
\begin{tabular}{c|c|c|c|c|c}
\hline
\multirow{2}{*}{ } 	& \multirow{2}{*}{Size Range} 			& \multirow{2}{*}{Crater Radius (m)} 	& \multicolumn{3}{c}{Ejecta Composition } 					\\
\cline{4-6}
                                 	&         							&                    	 				& Subrange				& Mass (kg) 			& Number	\\ 
\hline
\multirow{5}{*}{WCB}	& \multirow{5}{*}{$10\ \mu$m--$10$ cm} 	& \multirow{5}{*}{$3.4298$}  		& $10\ \mu$m--$100\ \mu$m 	& $3.4482 \times 10^3$	& $2.3617 \times 10^{14}$ \\
                                 	&          							&                              				& $100\ \mu$m--$1$ mm 		& $5.4650 \times 10^3$ 	& $3.7430 \times 10^{11}$ \\
                                 	&                                  				&                              				& $1$ mm--$1$ cm 			& $8.6615 \times 10^3$	& $5.9322 \times 10^8$ \\
                                 	&                                  				&                              				& $1$ cm--$10$ cm 			& $1.3727 \times 10^4$ 	& $9.4020 \times 10^5$ \\                                 
\cline{4-6}
                                 	&                                  				&                              				& Total 					& $3.1302 \times 10^4$ 	& $2.3654 \times 10^{14}$ \\                                 

\hline
\multirow{3}{*}{SFA}	& \multirow{3}{*}{$10\ \mu$m--$1$ mm} 	& \multirow{3}{*}{$7.1781$}  		& $10\ \mu$m--$100\ \mu$m 	& $6.4350 \times 10^4$	& $4.4074 \times 10^{15}$ \\
                                 	&          							&                              				& $100\ \mu$m--$1$ mm 		& $1.0199 \times 10^5$ 	& $6.9852 \times 10^{12}$ \\                                 
\cline{4-6}
                                 	&                                  				&                              				& Total 					& $1.6634 \times 10^5$ 	& $4.4144 \times 10^{15}$ \\                                 
\hline
\end{tabular}
\end{table}

\clearpage

\begin{table}[!ht]
\centering
\caption[]{Sampling parameters over the size subranges for the two considered materials and for ejection speeds either below or above $v_{cr}$. $N_s$ denotes the sample size (number of particles), and $w_s$ measures the weight of the sample's mass over the total mass of the considered subrange. }
\label{t:sampara2}
\vspace{0.2in}
\begin{threeparttable}
\begin{tabular}{c|c|c|c|c|c}
\hline
\multirow{2}{*}{ } 	& \multirow{2}{*}{Subrange} 	& \multicolumn{2}{c|}{$v \le v_{cr}$} 			& \multicolumn{2}{c}{$v>v_{cr}$} 	\\
\cline{3-6}
                                 	&         					&  $N_s$  		& $w_s$     				& $N_s$ 		& $w_s$ 	\\ 
\hline
\multirow{4}{*}{WCB}	& $10\ \mu$m--$100\ \mu$m 	&  $100,000$  	& $4.8149 \times 10^{-7}$ 	& $10,000$  	& $4.1107 \times 10^{-11}$ 	\\
				& $100\ \mu$m--$1$ mm 		&  $100,000$  	& $2.9632 \times 10^{-4}$ 	& $10,000$  	& $2.6381 \times 10^{-8}$ 	\\
				& $1$ mm--$1$ cm 			&  $100,000$  	& $1.9157 \times 10^{-1}$ 	& $10,000$  	& $1.6010 \times 10^{-5}$ 	\\				
				& $1$ cm--$10$ cm 			&  $800$  		& $9.1142 \times 10^{-1}$ 	& $10,000$  	& $1.0557 \times 10^{-2}$ 	\\
\hline
\multirow{2}{*}{SFA}	& $10\ \mu$m--$100\ \mu$m 	&  $200,000$  	& $2.4011 \times 10^{-9}$ 	& $20,000$  	& $4.3390 \times 10^{-12}$ 	\\
				& $100\ \mu$m--$1$ mm 		&  $200,000$  	& $1.5134 \times 10^{-6}$ 	& $20,000$  	& $2.8460 \times 10^{-9}$ 	\\
\hline
\end{tabular}
\begin{tablenotes}
\item[*] Note the critical radial distance, denoted by $x_{cr}$, is different for the two materials. For WCB, $x_{cr} = 3.4288$ m (close to the rim of the crater); and for SFA, $x_{cr} = 7.1324$ m. 
\end{tablenotes}
\end{threeparttable}
\end{table}

\clearpage

\begin{table}[!ht]
\centering
\caption[]{The terminal states of the ejecta cloud based on the results of the $12$ full-scale simulations. The mass of ejecta accreted on the primary, re-accreted on the secondary and escaped are abbreviated as M.A.P., M.R.S., and M.E., respectively. }
\label{t:fullres}
\vspace{0.2in}
\begin{tabular}{c | c | c c c}
\hline
							& $T_f$ (day) 	& M.A.P. (kg) 				& M.R.S. (kg) 				& M.E. (kg) 			 \\
\hline
 WCB $(0^\circ, 0^\circ)$			& $14.1286$    	& $6.0809 \times 10^{-1}$      	& $8.2803 \times 10^{-3}$		& $3.1302 \times 10^{4}$ 	\\
 WCB $(90^\circ W, 0^\circ)$		& $16.5485$    	& $4.3722 \times 10^{0}$      	& $1.7749 \times 10^{-1}$		& $3.1298 \times 10^{4}$ 	\\
 WCB $(180^\circ, 0^\circ)$		& $15.8214$    	& $4.2454 \times 10^{0}$      	& $3.7802 \times 10^{-2}$		& $3.1298 \times 10^{4}$ 	\\
 WCB $(90^\circ E, 0^\circ)$		& $8.4276$    	& $6.2952 \times 10^{-2}$      	& $1.0378 \times 10^{-2}$		& $3.1302 \times 10^{4}$ 	\\
 WCB $(0^\circ, 90^\circ N)$		& $23.9800$    	& $7.9702 \times 10^{-1}$      	& $1.6966 \times 10^{-1}$		& $3.1301 \times 10^{4}$ 	\\
 WCB $(0^\circ, 90^\circ S)$		& $28.5763$    	& $8.9824 \times 10^{-1}$      	& $3.0354 \times 10^{-2}$		& $3.1301 \times 10^{4}$ 	\\
\hline
 SFA $(0^\circ, 0^\circ)$			& $3.0484$    	& $1.2908 \times 10^{1}$      	& $1.2099 \times 10^{0}$		& $1.6632 \times 10^{5}$ 	\\
 SFA $(90^\circ W, 0^\circ)$		& $2.4126$    	& $4.0297 \times 10^{2}$      	& $2.4961 \times 10^{0}$		& $1.6593 \times 10^{5}$ 	\\
 SFA $(180^\circ, 0^\circ)$			& $1.9699$    	& $4.9881 \times 10^{0}$      	& $2.8210 \times 10^{0}$		& $1.6584 \times 10^{5}$ 	\\
 SFA $(90^\circ E, 0^\circ)$		& $1.9327$    	& $4.3341 \times 10^{0}$      	& $2.2397 \times 10^{0}$		& $1.6633 \times 10^{5}$ 	\\
 SFA $(0^\circ, 90^\circ N)$	   	& $2.2061$    	& $7.4368 \times 10^{1}$      	& $2.6276 \times 10^{0}$		& $1.6626 \times 10^{5}$ 	\\
 SFA $(0^\circ, 90^\circ S)$		& $2.9187$    	& $7.4703 \times 10^{1}$      	& $2.4478 \times 10^{0}$		& $1.6626 \times 10^{5}$ 	\\
\hline
\end{tabular}
\end{table}

\end{document}